\documentclass[aip,jap,reprint,tightenlines]{revtex4-1}

\usepackage{graphicx}
\usepackage{rotating}
\usepackage{amssymb}
\usepackage{amsmath}
\usepackage{breqn}
\usepackage{mathptmx}
\usepackage[colorinlistoftodos]{todonotes}
\usepackage{comment}
\usepackage{url}
\usepackage{etoolbox}

\makeatletter
\let\cat@comma@active\@empty
\makeatother

\newcommand{\apjs}{Astrophys. J. Supp.}

\newcommand{\nat}{Nature}

\begin{document}

\title{Horn-Coupled, Commercially-Fabricated Aluminum Lumped-Element Kinetic Inductance Detectors for Millimeter Wavelengths}

\author{H.~McCarrick}
\thanks{Electronic mail: hlm2124@columbia.edu}
\author{D.~Flanigan}
\author{G.~Jones}
\author{B.~R.~Johnson}
\affiliation{Department of Physics, Columbia University, New York, NY, 10025, USA}

\author{P.~Ade}
\affiliation{School of Physics and Astronomy, Cardiff University, Cardiff, Wales, CF24 3AA, UK}

\author{D.~Araujo}
\affiliation{Department of Physics, Columbia University, New York, NY, 10025, USA}

\author{K.~Bradford}
\affiliation{Department of Physics, Arizona State University, Tempe, AZ, 85287, USA}

\author{R.~Cantor}
\affiliation{STAR Cryoelectronics, Santa Fe, NM, 87508, USA}

\author{G.~Che}
\affiliation{Department of Physics, Arizona State University, Tempe, AZ, 85287, USA}

\author{P.~Day}
\affiliation{Jet Propulsion Laboratory, Caltech, Pasadena, CA, 91109, USA}

\author{S.~Doyle}
\affiliation{School of Physics and Astronomy, Cardiff University, Cardiff, Wales, CF24 3AA, UK}

\author{H.~Leduc}
\affiliation{Jet Propulsion Laboratory, Caltech, Pasadena, CA, 91109, USA}

\author{M.~Limon}

\author{V.~Luu}
\affiliation{Department of Physics, Columbia University, New York, NY, 10025, USA}

\author{P.~Mauskopf}
\affiliation{Department of Physics and School of Earth and Space Exploration, Arizona State University, Tempe, AZ, 85287, USA}
\affiliation{School of Physics and Astronomy, Cardiff University, Cardiff, Wales, CF24 3AA, UK}

\author{A.~Miller}
\affiliation{Department of Physics, Columbia University, New York, NY, 10025, USA}

\author{T.~Mroczkowski}
\thanks{National Research Council Fellow}
\affiliation{Naval Research Laboratory, Washington, DC, 20375, USA}

\author{C.~Tucker}
\affiliation{School of Physics and Astronomy, Cardiff University, Cardiff, Wales, CF24 3AA, UK}

\author{J.~Zmuidzinas}
\affiliation{Department of Physics, Caltech, Pasadena, CA, 91125, USA}
\affiliation{Jet Propulsion Laboratory, Caltech, Pasadena, CA, 91109, USA}
\date{\today}


\begin{abstract}

We discuss the design, fabrication, and testing of prototype horn-coupled, lumped-element kinetic inductance detectors (LEKIDs) designed for cosmic microwave background (CMB) studies.
The LEKIDs are made from a thin aluminum film deposited on a silicon wafer and patterned using standard photolithographic techniques at STAR~Cryoelectronics, a commercial device foundry.
We fabricated twenty-element arrays, optimized for a spectral band centered on 150~GHz, to test the sensitivity and yield of the devices as well as the multiplexing scheme.
We characterized the detectors in two configurations.
First, the detectors were tested in a dark environment with the horn apertures covered, and second, the horn apertures were pointed towards a beam-filling cryogenic blackbody load. 
These tests show that the multiplexing scheme is robust and scalable, the yield across multiple LEKID arrays is 91\%, and the measured noise-equivalent temperatures (NET) for a 4~K optical load are in the range 26$\thinspace\pm6$~$\mu \mbox{K} \sqrt{\mbox{s}}$.

\end{abstract}


\maketitle


\section{Introduction}
\label{sec:introduction}

In this paper we present the design and measured performance of horn-coupled, aluminum lumped-element kinetic inductance detectors (LEKIDs).
These devices were designed for cosmic microwave background (CMB) studies,~\cite{skip2013,Araujo2014} so they operate in a spectral band centered on 150~GHz, which is where the CMB frequency spectrum peaks.
Our LEKID design is scalable to higher frequencies, so these devices could be used for a range of millimeter-wave and sub-millimeter-wave activities.
The detectors were fabricated in industry, which is a unique aspect of this study.
To date, millimeter-wave detectors for CMB studies have exclusively been fabricated in government laboratories or at universities.
Here, we report the performance of the first generation of our commercially-fabricated devices.

LEKIDs are superconducting thin-film resonators also designed to be photon absorbers.
Absorbed photons with energies greater than the superconducting gap break Cooper pairs, changing the density of quasiparticles.
The quasiparticle density affects the kinetic inductance and dissipation of the superconducting film, so a changing optical signal will cause the resonant frequency and internal quality factor of the resonator to shift.
These changes in the properties of the resonator can be detected as changes in the amplitude and phase of a probe tone that drives the resonator at its resonant frequency.
This detector technology is particularly well-suited for sub-kelvin, kilo-pixel detector arrays because each detector element can be dimensioned to have a unique resonant frequency, and the probe tones for hundreds to thousands of detectors can be carried into and out of the cryostat on a single pair of coaxial cables.

Current experiments focused on studying the CMB polarization anisotropies use arrays of thousands of detectors.
Transition edge sensor (TES) bolometers are the current detector standard for these studies.
In general, TES devices are composed of a photon absorber suspended by a weak thermal link.
The absorber temperature is related to the incident photon power, and this temperature is detected with a superconducting temperature sensor and read out with a superconducting quantum interference device (SQUID).
The operational details of these detectors are thoroughly described in the literature.~\cite{Irwin2005}
A variety of TES architectures have already been deployed for CMB studies.~\cite{polarbear12,bicep2,sptpol,ebex,apexsz,actpol}
The next generation of CMB experiments will require an even greater number of detectors for improved sensitivity.
The inherent scalability of LEKIDs makes them a potential candidate for these future CMB measurements, so we conducted the study we describe here to explore this hypothesis further.
%


\begin{figure*}[!ht]
\centering
\includegraphics[width=.9\textwidth]{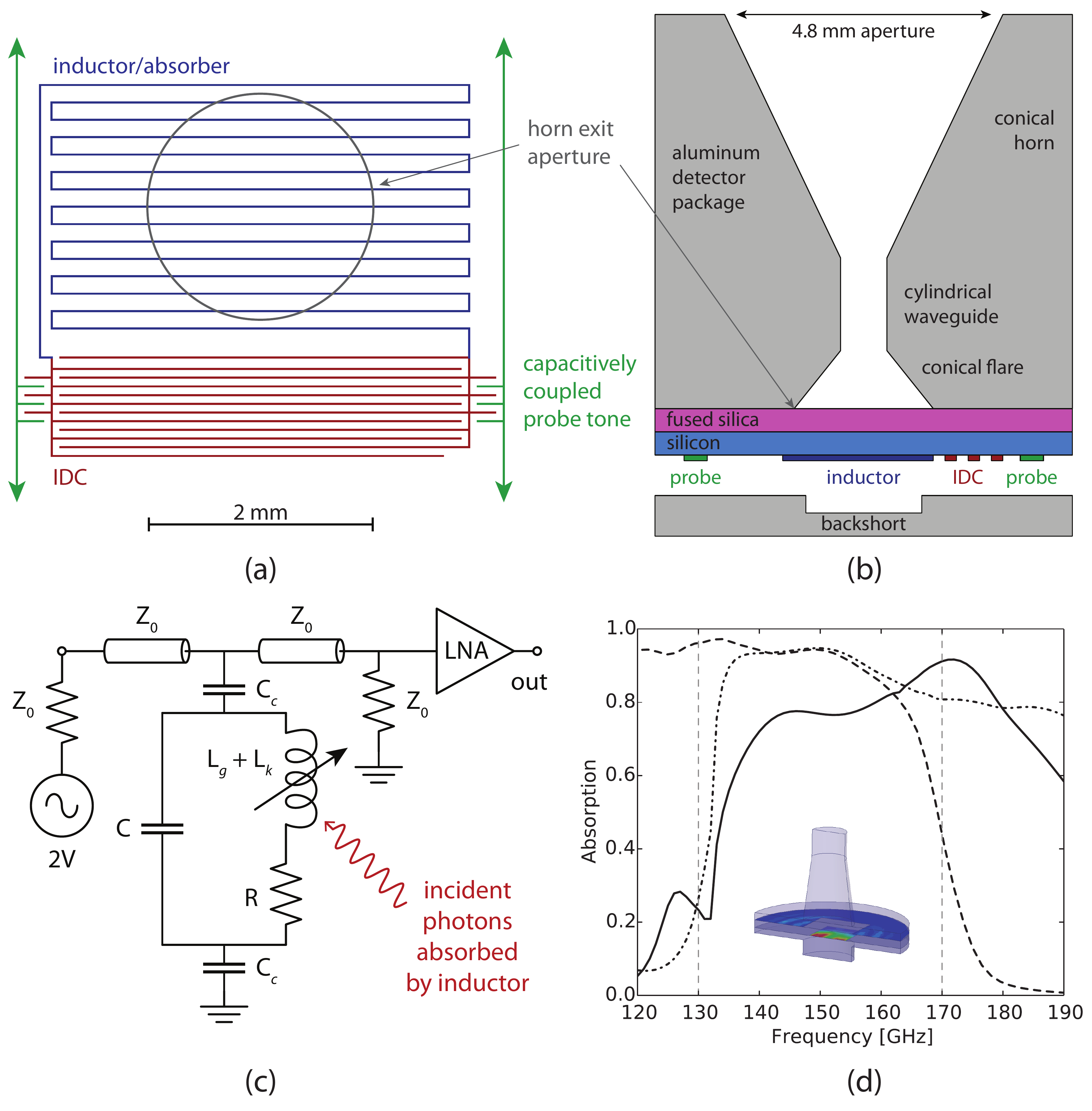}
\caption{{
\textbf{(a)} A schematic of a single LEKID.  The meandered inductor in the resonator is also the photon absorbing element. The interdigitated capacitor (IDC) completes the resonant circuit. The resonator is capacitively coupled to a transmission line, which carries the probe tone that is used to read out the detector. 
\textbf{(b)} Cross-sectional view of a single array element showing the horn plate, the dielectric stack, one LEKID, and the back-short plate.  For clarity, this schematic is not drawn to scale.
\textbf{(c)} Diagram of a single resonator circuit. The resonator is capacitively coupled ($C_{c}$) to the transmission line with impedance $Z_{0}$. The resonant frequency is set by the total inductance $L = L_{g} + L_{k}$, where $L_g$ is the geometric inductance and $L_k$ is the kinetic inductance, along with the capacitance $C$ of the interdigitated capacitor. An absorbed photon changes the quasiparticle density in the device and thus $L_{k}$, resulting in a resonant frequency shift, which is read out using a probe tone near the resonant frequency.
\textbf{(d)} Electromagnetic simulation results show the absorptance as a function of frequency for the horn-coupled aluminum LEKID design. The dot-dash line shows the single-polarization absorptance spectrum for the inductor/absorber using the nominal design value of 1.2~$\Omega/\square$ resistivity for aluminum.  The solid line shows the simulated absorptance spectrum for the measured resistivity of the devices, 4~$\Omega/\square$, which has an average absorptance of 72\% across the single-mode spectral band. The dashed line is the measured transmittance spectrum of the metal-mesh low-pass filter, used to define the upper edge of the spectral band. The inset shows the simulation set up and the simulated current density over the absorbing area.
}}
\label{fig:combined}
\end{figure*}


Microwave kinetic inductance detectors (MKIDs) were first published in 2003,~\cite{day} and the lumped-element MKID variety was published in 2008.~\cite{doyle}
Over the past decade, a number of groups around the world have pursued MKID technologies for a variety of astrophysical studies at different wavelengths, and our work builds from this experience.
Experiments that have deployed or plan to use MKID-based cameras include
ARCONS,~\cite{Mazin2013}
MAKO,~\cite{McKenney2012}
MicroSpec,~\cite{microspec2013}
MUSIC,~\cite{golwala+12}
NIKA,~\cite{monfardini}
BLAST-TNG,~\cite{dober2014}, and
SuperSpec.~\cite{superspec2012}
Laboratory studies show that state-of-the-art LEKID designs can achieve photon noise limited performance,~\cite{mauskopf14, McKenney2012} and photon noise limited horn-coupled LEKIDs sensitive to 1.2~THz were recently demonstrated.~\cite{Hubmayr2014}

The fundamental detector performance goal for CMB studies is to reduce the intrinsic detector noise so that it is negligible when compared with the noise due to the arrival statistics of the photon background. The detector performance reported in this paper is consistent with photon noise limited performance. 
LEKID noise has been extensively studied, and it includes contributions from three sources: generation-recombination (g-r) noise, amplifier noise, and two-level system (TLS) noise.
These noise sources are thoroughly described in the literature.~\cite{zmu}
The generation-recombination noise is due to fluctuations in the quasiparticle number from recombination into Cooper pairs and from thermal generation of quasiparticles.
Under typical loading conditions, this noise is caused mostly by randomness in the recombination of optically excited quasiparticles, and the thermal generation of quasiparticles is negligible.
The amplifier noise is the electronic noise of the readout system referred to the detector array.
It is set by the noise figure of the cryogenic microwave low-noise amplifier (LNA) immediately following the detectors.
TLS noise is produced by dielectric fluctuations due to quantum two-level systems in amorphous materials near the resonators.
The scaling of TLS noise with the operating temperature, probe tone power, resonant frequency, and geometry of the capacitor has been extensively studied experimentally.~\cite{Gao2008b}
This knowledge has been captured in a semi-empirical noise model, which gives us the ability to make empirically-grounded predictions of the TLS noise we should expect for a given capacitor and inductor design.~\cite{Gao2008b}
We designed our detectors using this semi-empirical model for the range of optical loads that are typical for ground-based and sub-orbital CMB experiments.

The remainder of the paper is organized in the following way.
In Section~\ref{sec:methods} we present the design of the horns and the detectors.
This section also describes the experimental set up, including the cryogenic system and the detector readout.
In Section~\ref{sec:results}, we present measurements of the LEKIDs with and without optical loading.
In Section~\ref{sec:conclusion}, we summarize our design and measurement results and describe our future plans. 
One goal of this paper is to provide a detailed end-to-end description of our design and testing process, which could be useful for uninitiated readers or groups interested in starting to make LEKIDs.
For clarity, in many places we provide the equations and other bits of practical information collected from the literature that were essential to our design and analysis process.


\begin{figure*}[t]
\centering
\includegraphics[width = .9\textwidth]{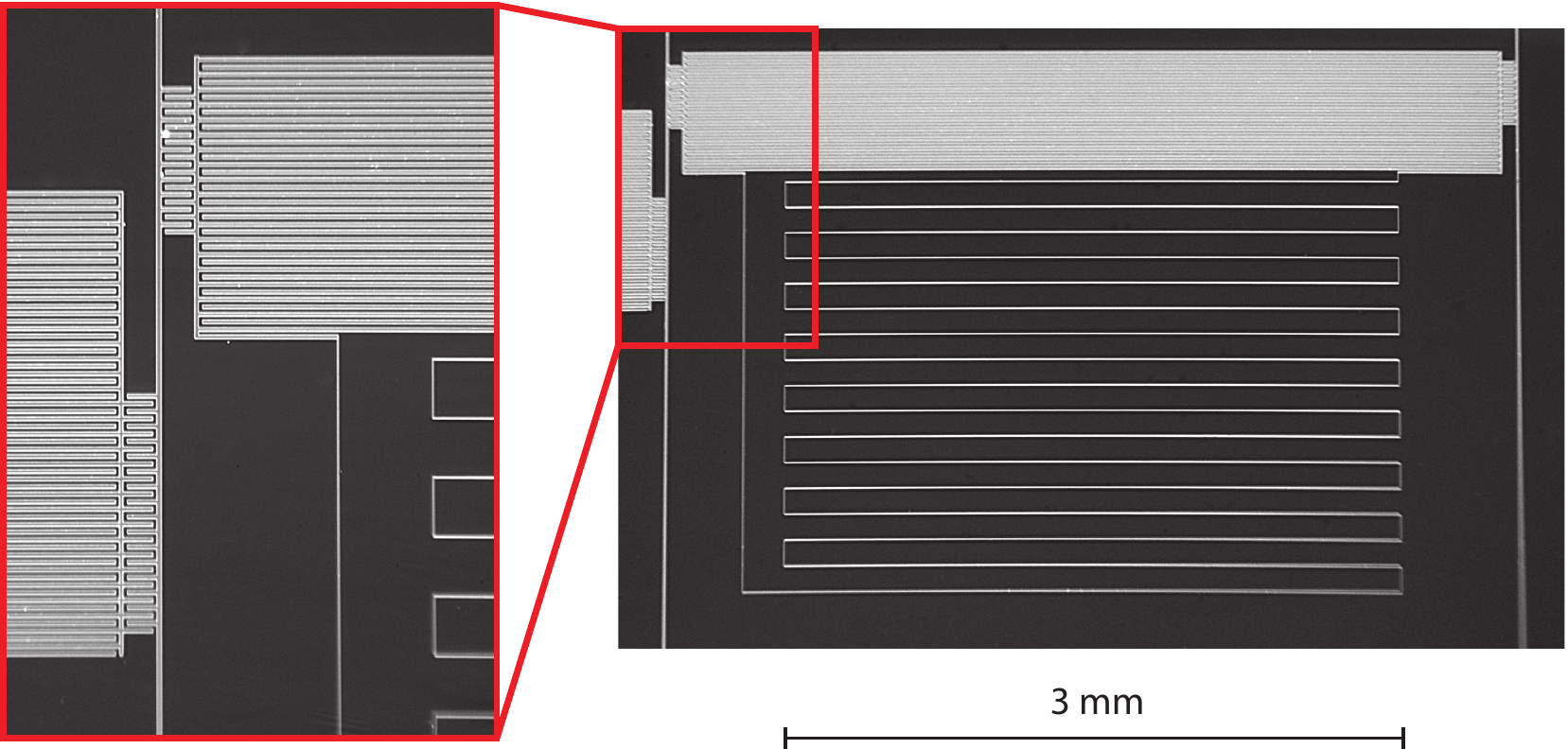}-eps-converted-to.pdf
\caption{{Photomicrographs of a single LEKID. All components are fabricated from thin-film aluminum with a single mask.}}
\label{fig:single_detector}
\end{figure*}


\section{Methods}
\label{sec:methods}

We designed and built a prototype twenty-element, horn-coupled LEKID module that is sensitive to a spectral band centered on 150~GHz. 
The module consists of a LEKID array on a silicon chip and an aluminum horn package.
The LEKIDs were fabricated in the foundry at STAR~Cryoelectronics in New Mexico. 
The aluminum horn package was manufactured in the Micromachining Laboratory at Arizona State University.
The modules were designed, assembled and tested at Columbia University.
For LEKIDs, the various construction parameters must simultaneously satisfy both the requirements of the resonator circuit and the optical coupling to millimeter-wavelength radiation.
It is instructive to first describe the photon coupling design (Section~\ref{sec:optical_coupling}) and then describe the resonator circuit (Section~\ref{sec:detector_design}).


\subsection{Optical Coupling Design}
\label{sec:optical_coupling}

Our design uses horn-coupled detectors for a number of reasons. 
First, the horn beam reduces sensitivity to stray light inside the cryostat and couples well with the telescope optics in instruments we are developing.~\cite{skip2013,Araujo2014}
Second, the waveguide in the horn provides an integrated high-pass filter.
Third, the horn pitch creates space for the large interdigitated capacitor, which allows for resonant frequencies below 250~MHz and reduces the effects of TLS noise.
%
%
Finally, electrical cross-talk is reduced because the final configuration is not tightly packed.

\begin{figure*}[t]
\centering
\includegraphics[width=.9\textwidth]{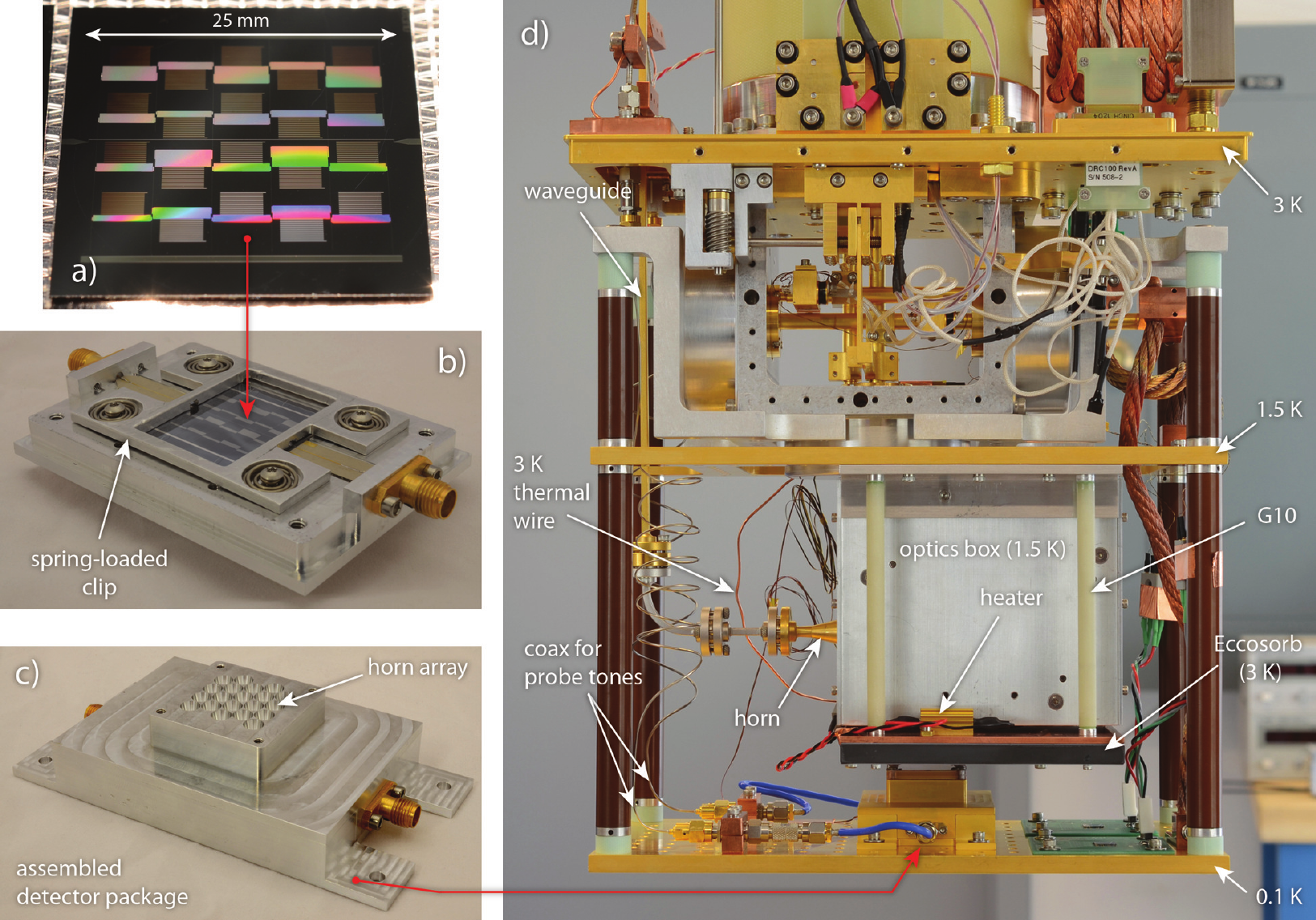}
\caption{{
\textbf{(a)} Photograph of a 20-element LEKID array. The LEKID inductors have a 4.8~mm hexagonal pitch, and the varying sizes of the interdigitated capacitors are evident. 
%
\textbf{(b)} The LEKID array mounted to the aluminum horn plate with the spring-loaded clip.  The horn apertures are facing down and therefore not visible in this photograph.
\textbf{(c)} The fully-assembled detector package with the conical horns facing up. For clarity, the low-pass filter, which is normally attached directly to the horn array using the four visible tapped holes, was removed for this photograph.
\textbf{(d)} The cryogenic test setup. The detector package is mounted to the 100~mK stage.  The Eccosorb is the optical load for this study, and the Eccosorb temperature was adjusted with the heater resistor.  The load is mechanically mounted to the 1.5~K stage with thermally isolating G10 legs and thermally connected to the 3~K stage with a copper wire, so the base temperature of the optical load should closely track the temperature of the 3~K stage.  The indicated waveguide, horn and optics box are elements of a second kind of load, which is under development and will be used in future studies. 
}}
\label{fig:experiment}
\end{figure*}

A cross-sectional view of one array element and a schematic of one detector are shown in Figure~\ref{fig:combined}, a photograph of a device is shown in Figure~\ref{fig:single_detector}, and the detector module is shown in Figure~\ref{fig:experiment}.  
The conical horn flare narrows down to a single-mode cylindrical waveguide section, which defines the low-frequency edge of the spectral band at 127~GHz.
The high-frequency edge is defined by a quasi-optical metal-mesh low-pass filter. 
The waveguide is then re-expanded with a second conical flare to reduce the wave impedance at the low-frequency edge of the spectral band, which improves optical coupling and allows the radiation to be launched efficiently into the subsequent dielectric stack.
The dielectric stack is composed of an approximately quarter-wavelength layer of fused silica (300~$\mu$m) and the silicon wafer (300~$\mu$m). The fused silica helps match the wave impedance to the silicon substrate.  
The radiation launched from the waveguide remains fairly well collimated as it travels through the dielectric stack and back-illuminates the inductor/absorber, which is patterned on the silicon and dimensioned to match the wave impedance.
There is a metal cavity behind each detector that is a quarter wavelength deep that acts as a back-short. 
The dielectric stack is mounted directly to the horn plate using a spring-loaded aluminum clip. 
The aluminum clip provides force that increases the thermal conductivity at the interface between the dielectric stack and the horn plate. 
A metal back plate, with the back-short cavities, is attached to seal the module.  
Electromagnetic simulations using the Ansoft HFSS software package predict that the maximum coupling efficiency to a single-polarization for this design is $90$\% and averages $>70$\% over the 130 to 170~GHz band, as shown in Figure~\ref{fig:combined}.

The conical flare produces a very small mixing of electromagnetic modes from the small aperture TE11 to the
exit aperture into the quartz of less than 1\%. There will be some reflection at the exit aperture and at the interface between the quartz and silicon and also at the detector. The electromagnetic simulations of the coupling to the detector, shown in Figure~\ref{fig:combined}, take all of these effects into account by launching a single mode (or one for each polarization) from the single mode circular waveguide section and then computing the electric field through the rest of the structure including the conical flare, the dielectric layers, the aluminum LEKID and the backshort. No resonances are seen in the band where the detector operates in part due to absorption of power by the aluminum and damping of any resonances. The size of the detector is large enough that it can effectively absorb multiple modes of incident electromagnetic radiation.


\subsection{Detector Design}
\label{sec:detector_design}

The detectors are designed to have the following properties: high absorptance and responsivity, low detector noise, and optimal electrical coupling.
The overall design of the detector requires balancing competing constraints on various construction parameters. 
In the following paragraphs we describe the design of the inductor, which largely controls the absorptance and responsivity, the capacitor geometry, which influences the TLS noise, the transmission line, and the electrical coupling. 

The detector array consists of back-illuminated LEKIDs  fabricated from a 20~nm thick aluminum film deposited on a 300~$\mu$m thick high-resistivity ($>10$~k$\Omega$cm), float-zone silicon substrate.
The inductor/absorber is a meandered aluminum trace on silicon with a filling factor of 1.5\%, calculated as the inductor trace width divided by the gap width plus trace width. 
This filling factor is designed to match the wave impedance of the incoming radiation in silicon, which has a dielectric constant, $\epsilon_r=$~11.9 and a wave impedance of $\sim$110~$\Omega$. 
The effective impedance of the inductor is $Z_{eff} \approx (\rho\,g_L)/(w_L\,t)$, where $t$ is the film thickness, $g_L$ is the gap width between meanders, $w_L$ the meander width, and $\rho$ the material resistivity, for which we used the typical value for 20~nm thick aluminum, 1.2~$\Omega/\square$.

To efficiently absorb incident photons and approximate a solid sheet, the pitch between the inductor meanders should be less than $\lambda$/20 where $\lambda$ is the incident wavelength.~\cite{Doyle1} 
The choice of a 2~$\mu$m wide trace and 125~$\mu$m spacing between meanders, gives an effective sheet impedance of 76~$\Omega$ with a 15\% efficiency loss, using the standard value for 20~nm thick aluminum resistivity.
Although this does not perfectly match the wave impedance in the silicon, the absorption is not particularly sensitive to this parameter. 
In Figure~\ref{fig:combined}, electromagnetic simulations using the HFSS software package show the detector absorption  using both the nominal design resistivity and the measured resistivity of the devices, 4~$\Omega/\square$. 
Electromagnetic simulations also show that the incident radiation spreads preferentially in the direction perpendicular to the E-field and thus the inductor/absorber has dimensions of $2\times 3$~mm. 

In addition to absorbing the incident photons, the inductor is also part of the resonator circuit. 
Thus, the meandered inductor must be carefully designed to match the wave impedance of the incident photons yet have a high kinetic inductance fraction, $\alpha_k=L_k/(L_k+L_g)$.
The kinetic inductance can be predicted using
\begin{equation}
L_k = \frac{A_L}{w_L(w_L + g_L)} \frac{h R_s}{2 \pi^{2} \Delta},
\end{equation}
where $w_L$ is the width of the inductor trace, $A_L$ the total area of the inductor, $R_s$  the normal surface resistance, $h$ the Planck constant, and $\Delta$  the gap energy, defined as~\cite{Tinkham2004} $\Delta \approx 1.76 k_B T_c $. Here, $T_c$ is the superconducting transition temperature and $k_B$ is the Boltzmann constant.  The first term is simply the number of squares of material and the second term in the expression is the kinetic surface inductance $L_s$. The value for the geometric inductance was obtained from electromagnetic simulations using the Sonnet software package. Practical fabrication constraints with a contact mask limit the film thickness to approximately 20~nm and the trace width to approximately 2~$\mu$m. 
For a 20~nm thick by 2~$\mu$m wide aluminum trace, the predicted $L_k \approx 35$~nH and $\alpha_k \approx 0.4$.

Fluctuations in the absorbed optical power are proportional to fluctuations in the total quasiparticle number $N_{qp}$ and as such, fluctuations in the quasiparticle density $n_{qp} = N_{qp} / V_L$ are inversely proportional to inductor volume. 
Thus, for a given optical load, decreasing the volume of the inductor increases the responsivity.
Given the above constraints, the inductor volume is $V_{L} = 1870~\mu$m$^{3}$. 
The geometric inductance of the resulting design is approximately 59~nH. 

Changes in the quasiparticle density in the resonator cause changes in both the resonant frequency and the quality factor, which is related to the internal dissipation.
These changes cause deviations of the complex transmission in orthogonal directions, which are referred to as the frequency and dissipation directions. 
Two-level systems produce noise only in the frequency direction.~\cite{Gao2007,Gao2008a}

Suppressing the TLS noise has historically been challenging. 
Gao et al.~\cite{Gao2008b} have shown that TLS noise scales with the capacitor digit gap widths as $g_{IDC}^{-1.6}$, and thus we want to maximize the gap widths.
We also chose to target resonant frequencies below 200~MHz to match the baseband bandwidth of the readout system, avoiding the need for mixers.
Using a low readout frequency allows us to fit more detectors in a given bandwidth. 
%
Lower readout frequencies should in theory couple less strongly to the TLS fluctuators in the resonators, thus reducing their impact.~\cite{zmu}
To simplify modeling, we imposed the constraint on the interdigitated capacitors that the digit widths and the gaps between them be of equal size.~\cite{LimY} 
We maximized the gap widths, while maintaining a resonant frequency less than 200~MHz using the available area as determined by the horn and detector pitch.
The resulting capacitor geometry has a gap width of 8~$\mu$m. The area of the largest capacitor is 9~mm$^{2}$. The capacitors have values in the range 6 to 28~pF.


\begin{figure}[t]
\centering
\includegraphics[width=\columnwidth]{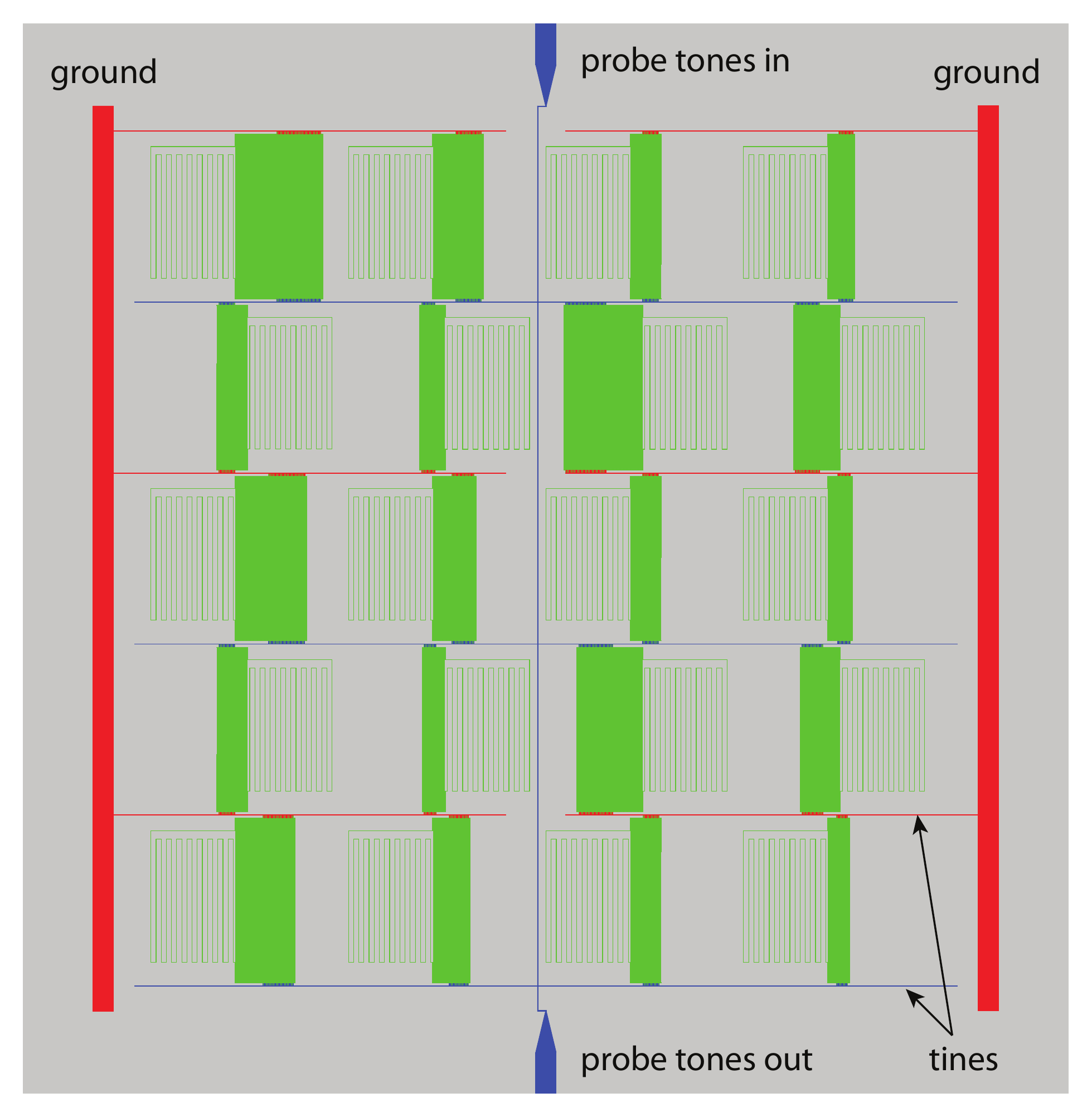}
\caption{The detectors are coupled to a feed line which runs down the center of the array and has tines that distribute the probe tones to the individual LEKIDs.}
\label{fig:tines}
\end{figure}


The resonators are capacitively coupled to the transmission line, which carries the probe tones.
We use an aluminum transmission line structure that has a central trace across the array and tines which distribute the signal to the individual detectors, as shown in Figure~\ref{fig:tines}. 
The transmission line structure acts as a lumped element as its length is much less than the wavelength of the readout frequencies. 
Ground returns for the resonators are provided by similar tines coming from aluminum strips at the sides of the chip, which are wire bonded to the package.
The central transmission line is a trace 20~$\mu$m wide.
The tines are 15~$\mu$m wide. 
We chose these widths based on simulations, which show that the transmission line is matched to 50~$\Omega$ across the readout bandwidth at the interfaces with the rest of the readout chain. 

To achieve sufficient coupling at these low resonant frequencies we used interdigitated capacitors between the resonator and both the signal tine and the ground return tine. 
This coupling design is schematically shown in Figure~\ref{fig:combined}. 
Electromagnetic simulations were used to verify this coupling scheme. 
To maximize responsivity, the coupling quality factor $Q_{c}$ should equal the internal quality factor $Q_{i}$ under the expected optical load.~\cite{zmu}
To calculate the necessary coupling capacitance we begin with the definition of the quality factor
\begin{equation}
Q \equiv \frac{2 \pi f_0 E}{P_d}, 
\end{equation}
where $E$ is the peak energy stored in the resonator, $f_0$ is the resonant frequency, and $P_d$ is the average power dissipation.
During the phase of the oscillation when the resonator energy is completely stored in the capacitor, we can write the coupling quality factor $Q_c$ as
\begin{equation}
\label{eqn:qc}
Q_{c} = \frac{2 \pi f_0}{P_d} \bigg(\frac{1}{2}C|V|^{2}\bigg), 
\end{equation}
where $P_d$ is the power dissipated from the resonator into the load impedance $Z_0$ across the coupling capacitor, $C$ is the capacitance of the main capacitor in the resonator, and $V$ is the peak voltage across the resonator.
The power dissipated in the load through the coupling capacitor is then
\begin{equation}
P_{d} = \frac{1}{2}|I|^{2}\frac{Z_0}{2},
\end{equation}
where $I$ is the current that flows through the coupling capacitor, $C_c$.
Since $1/(2 \pi f_0 C_c) \gg Z_0$, and taking into account the fact that there are two coupling capacitors in series, we can write
\begin{equation}
\label{eqn:powerdiss}
P_d = \frac{1}{2}\bigg|V \frac{2 \pi j f_0 C_c}{2}\bigg|^{2}\frac{Z_0}{2}.
\end{equation}
Finally, by substituting Equation~\ref{eqn:powerdiss} into Equation~\ref{eqn:qc}, we arrive at
\begin{equation}
C_{c} = \sqrt{\frac{8 C}{2 \pi f_0 Q_c Z_0 }},
\end{equation}
and for the optimal coupling of this design we set $Q_c = Q_i = 10^{5}$.
Thus, the coupling capacitors are designed to have $C_{c}$ values between 0.06 and 0.25~pF.

 
\subsection{Detector Fabrication}

The wafers were processed at STAR~Cryoelectronics using standard photolithographic procedures.
First, the wafers were cleaned with acetone, isopropyl alcohol, and deionized water. A plasma ashing was used to remove any residual organic material. 
An argon plasma was then used to remove SiO$_2$ from the surface of the wafer. 
The aluminum film was deposited through evaporation.
The wafers go through standard lithography: application of hexamethyldisilazane (HMDS) to promote resist adhesion, dehydration bake, resist coat, soft bake, resist exposure, resist developing, hard bake, descum and finally ion milling is used to etch away the unwanted aluminum. 
We used a standard contact mask for patterning the aluminum. 
The resist was then stripped using acetone, isopropyl alcohol, deionized water, and plasma ashing, put in vacuum to dehydrate, coated with HMDS, and finally a protective resist layer is applied to protect the devices while dicing.
Photomicrographs of the fabricated devices are shown in Figure~\ref{fig:single_detector}.

A common technique used for MKID fabrication on silicon is to dip the wafer in a hydrofluoric acid bath before processing. 
This etches away any SiO$_2$ and hydrogen-terminates the silicon, protecting it from further oxidation. 
This step can help reduce TLS effects. 
We have fabricated additional devices including this step, the results of which we will describe in a future paper.


\begin{figure*}[t]
\centering
\includegraphics[width=\textwidth]{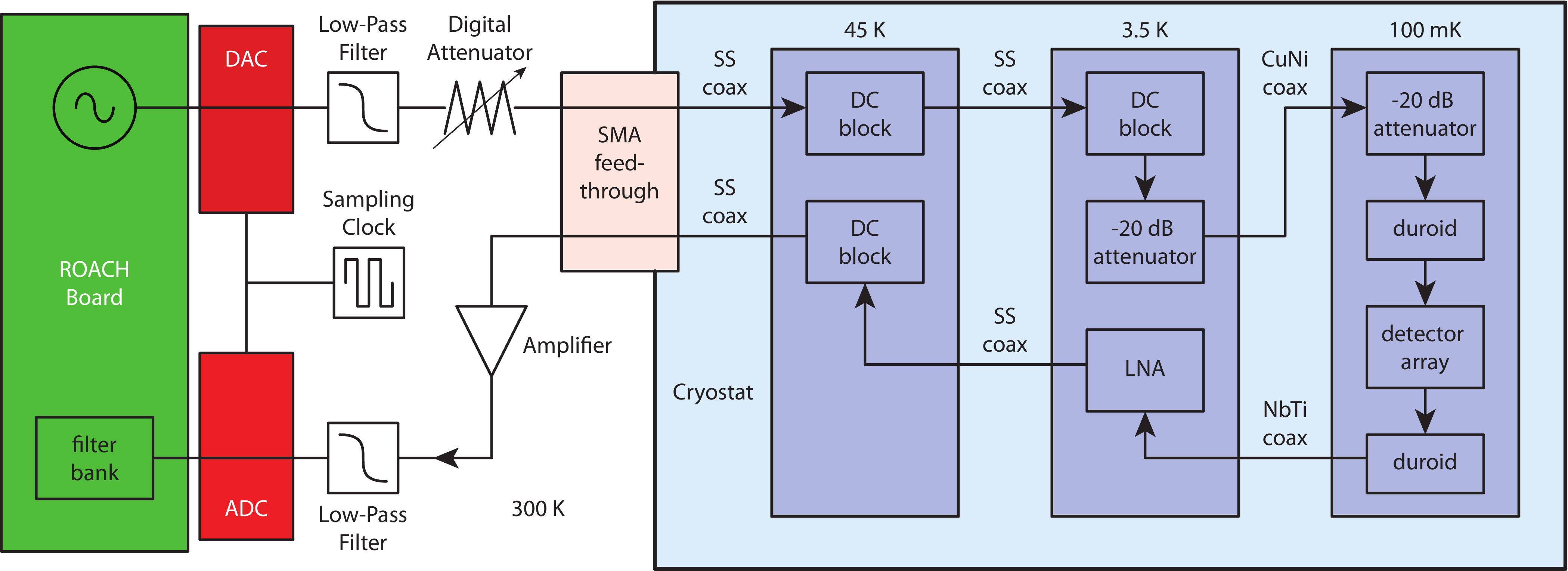}
\caption{A block diagram of the probe tone signal chain.}
\label{fig:circuits}
\end{figure*}


\subsection{Experimental Setup}

The detectors are cooled in a DRC-102 ADR Cryostat System made by STAR~Cryoelectronics.
This cryostat system uses a Cryomech PT407 Pulse Tube Cooler and a two-stage adiabatic demagnetization refrigerator (ADR), which provides 112~mJ of cooling capacity for the 100~mK stage.
The working end of the cryostat is shown without radiation shields in Figure~\ref{fig:experiment}.

All of the detectors in the module are frequency-multiplexed in a readout band between 80~and~160~MHz, and read out with a single SiGe bipolar cryogenic low-noise amplifier (LNA) and one pair of coaxial cables.~\cite{Weinreb2007} 
Here we describe in detail the path of the probe tones, shown schematically in Figure~\ref{fig:circuits}.
To minimize the loading from the coax on the 4~K stage, a 1~m long 2.16~mm (85~mil) diameter stainless steel inner and outer conductor coax (SSI Cable Corps UT085-SS) runs from ambient temperature to 4~K, intercepted at the 45 K stage. 
DC blocks (Inmet 8040) provide thermal breaks for the center conductor of the coax at the 45~K and 4~K stages.
After the DC block, a -20~dB attenuator (Inmet GAH-20) reduces any 300~K radiation propagating down the coax, dissipating the power at the 4~K stage.
A 60~cm long section of 0.86~mm (34~mil) diameter cupronickel (CuNi) coax (Coax Co. SC-033/50-CN-CN\footnote{\protect\url{http://www.coax.co.jp/english}}) carries the signal from 4~K to another -20~dB attenuator mounted on the 100~mK stage which reduces the noise contribution from the 4~K stage attenuator. We originally used an Anritsu 43KB-20 for the coldest attenuator, which had good performance, but started superconducting around 500~mK, rendering it unsuitable for this application.
A short length of semi-rigid copper coax brings the signal from a second attenuator on the 100~mK stage to the cooled package.
The copper coax terminates in an SMA connector on the package, which is soldered to a gold-plated microstrip interface board made from Duroid~6010 ($\epsilon_r=$ 10.8).
Aluminum wire bonds connect the microstrip to the detector array, and explicitly connect the ground return of the detectors to the package. 

On the return path, a 50~cm long superconducting NbTi coax (Coax Co. SC-033/50-NbTi-NbTi) carries the signal from 100~mK to 4~K, which has both extremely low thermal conductivity and low loss.~\cite{kushino2008}
The signal is then amplified by the LNA (Caltech LF-2), with a noise temperature, $T_N <$5~K, over the array bandwidth.
The gain of the LNA is sufficiently high that the noise temperature of the cascaded readout chain is dominated by that of the LNA, which itself is subdominant to the noise of the detector. 
Stainless steel coax carries the signal from 4~K to 45~K and finally to 300~K, with a DC block at the 45~K stage.


\subsubsection{Digital Readout}

Detector data was collected with a digital readout system called CUKIDS that was developed at Columbia University.
This system uses the CASPER signal processing tool flow, a ROACH field-programmable gate array (FPGA) board,\footnote{\protect\url{http://casper.berkeley.edu}} and the 12-bit analog-to-digital converter and 16-bit digital-to-analog converter (DAC) card developed for the MUSIC instrument.~\cite{duan+10}
The room temperature analog signal conditioning consists of amplifiers, low-pass filters, and digital step attenuators from Mini-Circuits, Inc.
The readout firmware, control, and analysis software we developed is open-source and available online.\footnote{\protect\url{https://github.com/ColumbiaCMB/kid_readout}}

Currently, the system is optimized for laboratory testing, essentially providing multiple homodyne test setups in parallel.
The probe tone waveforms are generated using a circular playback buffer feeding the DAC.
After digitization, the signal is channelized using a polyphase filterbank (PFB).
The complex voltage waveforms from the PFB channels that contain the probe tones are sent to the host computer for storage and analysis.
All subsequent demodulation and analysis is done offline.
For the measurements reported here, the FPGA was configured to provide four simultaneous homodyne readouts, each with a bandwidth of 125~kHz. 
%
%
\subsubsection{Blackbody Load}
For this study, we designed and built the cryogenic test setup shown in Figure~\ref{fig:experiment}.
This setup includes a blackbody load that has a temperature range from less than 4~K to 6~K. 
In its coldest state, it should look similar to the CMB.
The selected loading range is similar to those that are expected for ground-based and sub-orbital experiments. 
We use the variable temperature load to directly calibrate our detector noise and responsivity.

The blackbody is constructed from a slab of 6.35~mm (0.25~inch) thick Eccosorb MF-110 absorber coated with a 0.4~mm (0.015~inch) thick sheet of etched Teflon for impedance matching.
The etched Teflon is bonded to the Eccosorb with a thin layer of Stycast 2850FT.
We designed this load using loss tangent and refractive index information from the literature.~\cite{Peterson1984} The emissivity of the load is calculated to be 92\%. 

The Eccosorb slab is mounted to a copper thermal bus, and this assembly is mechanically supported by G-10 legs that are connected to the 1.5~K ADR stage.
The temperature of the Eccosorb is controlled using a weak thermal link to the 4~K pulse-tube cooler stage and a heater resistor, which is mounted on the copper thermal bus.
We designed the thermal time constant of the blackbody source to be approximately 20~minutes to minimize the required heat input while still providing a reasonably short settling time when changing temperatures.
The blackbody is \textless 1~cm from the detector module, and the entire setup is enclosed in a 4~K shield.
There is also a copper shield surrounding the detector module to minimize light leaks.
This shield is attached to the 100~mK stage, and it has an aperture exposing the low-pass filter.


\section{Results}
\label{sec:results}

We performed a range of experiments to measure the quality of the fabricated detectors, the results of which are presented in the following sections. We first report the electrical properties of the film, followed by measurements of the detectors themselves in a dark environment. We describe in detail the fitting procedure used to analyze the data. The data are compared to Mattis-Bardeen theory. We then proceed to optical testing with the blackbody load described above, which provides measurements of the optical responsivity and noise.


\subsection{Film Properties}
\label{sec:film_properties}

To ascertain the residual resistance ratio (RRR) and the superconducting transition temperature $T_c$ of the aluminum film, we performed a four-wire measurement of the resistance of a $2 \, \mu \mathrm{m} \times 35,000 \, \mu \mathrm{m}$ meandered trace as a function of temperature.
This witness sample was made alongside the LEKIDs on the same silicon wafer and therefore from the same 20~nm thick aluminum film.
The resistance of the sample at 3~K is 70~k$\Omega$, yielding a surface resistance of 4.0~$\Omega/\square$. 
The resistance at 300~K is approximately 210~k$\Omega$, giving a measured  RRR of 3.3.
We measured $T_c = 1.46 \, \mathrm{K}$, which agrees well with the $T_c$ measured independently by the probe tones at the $\sim$100~MHz readout frequencies.
Other measurements of thin-film aluminum in the literature~\cite{Meservey71} also report values of the critical temperature higher than that of bulk aluminum, which is nominally 1.2~K.


\subsection{Dark Testing}

Initial characterization of the resonators is done in a ``dark'' package, which is sealed with metal tape to minimize light leaks.
Frequency sweeps through the resonances taken at different bath temperatures can be fit to determine the resonant frequencies and quality factors as a function of temperature.
Frequency sweeps are done at a variety of probe tone powers to determine the maximum readout power at which the detector can be operated before the device response becomes non-linear due to the non-linear kinetic inductance effect.~\cite{Swenson2013}
The bifurcation power is found to be around -100~dBm across the array. 
%
%
This is approximately 10 dB higher than predicted using the theory described by Swenson et al.,~\cite{Swenson2013} assuming that the non-linearity energy scale $E_{*}$ is equal to the condensation energy of the inductor $E_{\mathrm{cond}} = N_0 \Delta ^2 V_L /2$, which is expected to be the case if $\alpha_k\approx 1$. 
Here $N_0$ is the single-spin density of states at the Fermi level.
Approximately 4 dB of this discrepancy can be explained by the fact that $\alpha_k\approx 0.66$. 
In addition, as Swenson et al. point out, it is difficult to directly compare $E_{\mathrm{cond}}$ to the value of $E_{*}$ implied by measurements because the absolute power in the inductor is influenced by unknown temperature dependent loss in the cryogenic cabling and by mismatches between the transmission line and the resonator.
The measurements reported here were taken with a readout power of approximately $-111$~dBm, well below bifurcation, except where otherwise noted.


\subsubsection{Yield}

Overall, we have tested a total of three 20-element arrays and two 9-element arrays all made on the same wafer. 
We have found 71 working resonators of the total 78, corresponding to an overall yield fraction of $\sim$91\%.
Subsequent test results focus on a single 20-element array. 
The resonant frequencies were designed to fall between 100 to 200~MHz.
We found that all were systematically shifted down in frequency by about 15\%.
This frequency shift is reasonably well explained when the kinetic inductance is calculated using the measured surface resistance of 4~$\Omega / \square$ and $T_c=1.46$~K instead of the originally assumed 1.2~$\Omega/\square$ and $T_c=1.2$~K, giving an $L_k$ of approximately 100~nH and a resonance frequency shift of approximately 20\%. 
Lithographic tolerances (e.g. the under etching of the capacitor resulting in increased capacitance or a different film thickness than desired) could also be responsible for smaller shifts in the resonant frequency.


\subsubsection{Resonator Frequency Sweep Fitting}

We fit the resonators using a model which takes into account the skew of the resonance caused by mismatches in the transmission line.~\cite{Khalil+2012}
At higher probe powers, we found it necessary to also incorporate the nonlinear model presented by Swenson et al.~\cite{Swenson2013}
The complete model for the complex forward transmission $S_{21}$ as a function of frequency $f$ is:
\begin{equation}
S_{21}(f) = A e^{-2\pi jDf} \left(1-\frac{Q_r/Q_e}{1+2jQ_rx}\right),
\label{eqn:resonator}
\end{equation}
where $A=|A|e^{j\phi}$ is an arbitrary complex scale factor, $D$ is the cable delay, $Q_r$ is the loaded resonator quality factor, $Q_e$ is the complex coupling quality factor, and $x$ is the detuning parameter, which is simply $(f-f_0)/f_0$ in the case of the basic linear model, where $f_0$ is the resonant frequency.
For the nonlinear bifurcation model, $x$ is given by the solution to the cubic equation
\begin{equation}
y = y_0 + \frac{a}{1+4y^2},
\end{equation}
where $y_0=(f-f_0)/f_0$, $x = y/Q_r$, and $a$ is the bifurcation parameter defined by Swenson et al.~\cite{Swenson2013} The solution to this cubic equation is given in the appendix.

The real and imaginary parts of the model and data were fit simultaneously using the Levenberg-Marquardt algorithm for non-linear least-squares minimization.
We parameterized $Q_e$ in terms of its real and imaginary components, which yielded more robust fits than using its magnitude and phase.
%
By fitting both the real and imaginary parts of the model simultaneously, we found that the resulting fits were very well constrained, even with only a few data points spaced across the resonance.
We used the {\tt lmfit} Python package\footnote{\protect\url{http://cars9.uchicago.edu/software/python/lmfit}} which provides a convenient interface to the underlying algorithm.
We also used the {\tt emcee} package~\cite{ForemanMackey2013} to perform a Markov-Chain Monte Carlo analysis of the fits to ensure the errors were realistic.


\subsubsection{Quality Factors} 

We adopt the convention proposed by Khalil et al.,~\cite{Khalil+2012} defining the internal (unloaded) quality factor of the resonator as
\begin{equation}
Q_i^{-1} = Q_r^{-1} - \operatorname{Re} Q_e^{-1}.
\end{equation}
We define an effective real coupling quality factor
$Q_c = \left( \operatorname{Re} (Q_e^{-1}) \right)^{-1}$. 

\begin{figure}[t]
\centering
\includegraphics[width=0.95\columnwidth]{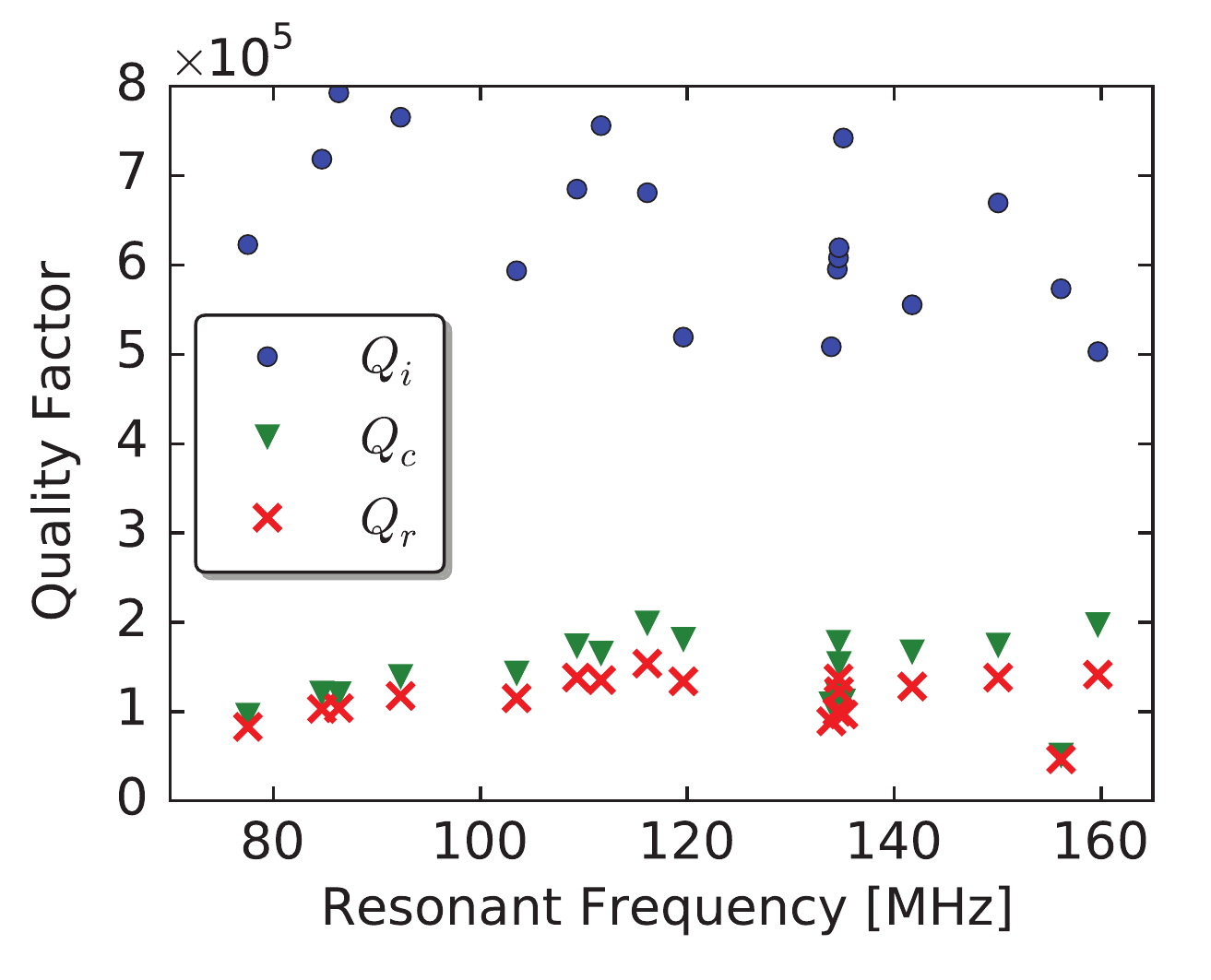}
\caption{{
Quality factors for the resonators measured at 200~mK in a dark environment. The internal quality factors are all greater than $5\times10^5$.
Here, $Q_c$ refers to $(\operatorname{Re} (Q_e^{-1}))^{-1}$.
The errors on $Q_r$ and $Q_c$ are typically $\sim1$\%, while the errors on $Q_i$ are around 10\%.
}}
\label{fig:quality_factors}
\label{fig:load_sweep}
\end{figure}

The quality factors for the detectors measured in an aluminum package and a dark environment at 200~mK are greater than $5\times10^5$ as shown in Figure~\ref{fig:quality_factors}.
The package is made of the QC-10 aluminum alloy,\footnote{\protect\url{http://www.alcoaqc10.com}} which is easily machinable and known to superconduct in our operating temperature range.
We had originally used a gold-plated copper package, but found that the internal quality factors were limited to $\sim 4 \times 10^4$, presumably due to coupling between the resonators and the lossy normal metal of the package.

The effective real coupling quality factors are $\sim 2 \times 10^{5}$ and match reasonably well to the design value. With no loading, the coupling quality factor limits the resonator quality factor.
Under optical loading, however, the coupling is better matched.   
The resonator quality factors are sufficiently high that $\sim$~300 resonators can be read out in a single octave~\cite{swenson1} as required for the proposed experiments.
Additionally, the resonant frequencies of five detectors on the tested array were spaced such that greater than 300 resonators could be read out in a single octave. 
These resonators were all consistently functional and did not collide.


\subsubsection{Bath Temperature Sweeps}
\label{sec:bath_sweeps}


\begin{figure}[t]
\centering
\includegraphics[width=\columnwidth]{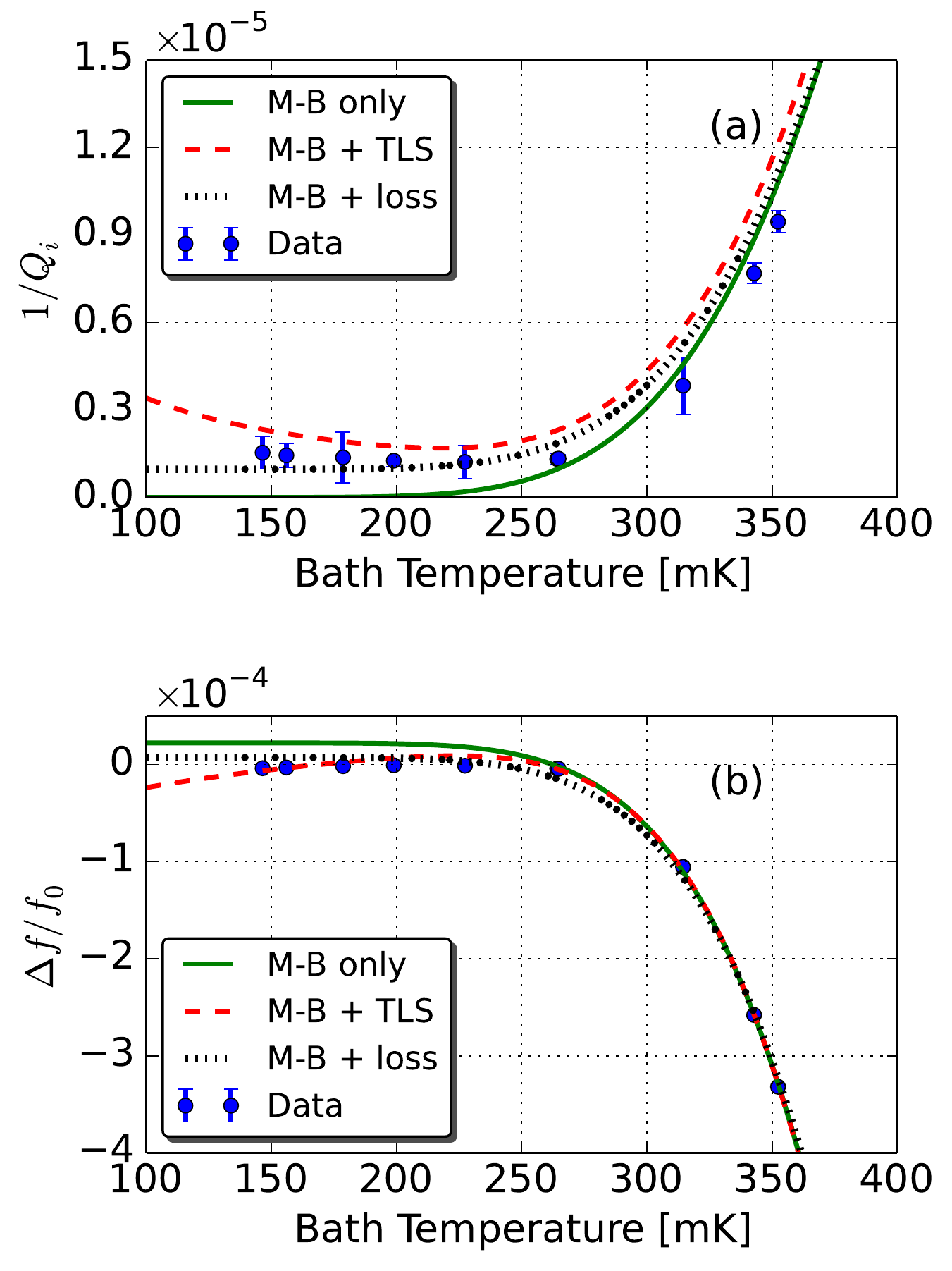}
\caption{{
Bath temperature sweeps for a single resonator. The resonant frequency at 200~mK is 86.31195~MHz, and the probe power used was $\sim-111$~dBm.
The plot in panel \textbf{(a)} shows the inverse internal quality factor, and
the plot in panel \textbf{(b)} shows measurements of the fractional frequency change.
%
%
%
%
%
%
%
Joint fits to the data for three models are plotted: Mattis-Bardeen theory alone (solid green), M-B with temperature-dependent TLS loss (dashed red), and M-B with a fixed loss term (dotted black).
The $x_{\mathrm{offset}}$ parameter is also included in the models plotted in panel \textbf{(b)}.
For all models, $\alpha_k$ was fixed at 0.65, while $T_c$ was allowed to vary. 
All fits resulted in $T_c\approx1.60$~K, implying that the relationship between $\Delta$ and $T_c$ is closer to $\Delta \approx 1.93 k_B T_c$, assuming $T_c$ is actually 1.46~K as measured in Section~\ref{sec:film_properties}.
The M-B theory provides a good fit to our data above 250~mK, particularly for the frequency response.
The fit to the TLS model yielded $F_{\mathrm{TLS}}\delta_0\approx 1.4 \times 10^{-4}$, which is driven by the ``back-bending" observed in the frequency response.
The shape of this curve is inadequate to explain our data, so we do not place much confidence in this model, and hence the resulting value of $F_{\mathrm{TLS}}\delta_0$ for our devices.
The model with constant loss term fits the data reasonably well (yielding $Q_{i,loss}^{-1}\approx 1.0 \times 10^{-6}$) but does not explain the ``back-bending" behavior.
}}
\label{fig:mb_fit}
\end{figure}


The bath temperature of the detectors $T$ is stepped in order to measure the device responsivity to thermal quasiparticles, to compare the device response to Mattis-Bardeen theory, and to look for signatures of TLS effects. 

At each temperature, a frequency sweep of each resonator was measured and fit to Equation \ref{eqn:resonator} to extract the resonant frequency and quality factors.
For each resonator, the fractional change in resonant frequency was computed as 
$x = (f-f_{max})/f_{max}$, where $f_{max}$ is the maximum observed value of the resonant frequency for a resonator for the experiment.

At each temperature step and for each resonator a simultaneous fit was performed to the following two equations:
\begin{equation}
Q_{i,total}^{-1} = Q_{i,\mathrm{MB}}^{-1} + Q_{\mathrm{TLS}}^{-1} + Q_{\mathrm{loss}}^{-1}
\end{equation}
\begin{equation}
x = x_{MB} + x_{\mathrm{TLS}} + x_{\mathrm{offset}}.
\end{equation}
The terms in the expression for $Q_{i,total}^{-1}$ are the prediction from Mattis-Bardeen theory, the loss due to TLS, given below, and a constant loss term to account for effects like radiation to free space or coupling to lossy materials near the device, which do not depend strongly on temperature.
%
%
%

In the following equations, we generally follow the treatment in Noroozian.~\cite{Noroozian2012b}
The equations assume that the film is thin, that $hf \ll \Delta$, and that $k_B T \ll \Delta$.
All of these assumptions are well satisfied for our devices. 
Using these assumptions, $Q_{i,MB}$ is given by
\begin{equation}
\label{eqn:qimb}
Q_{i,MB} = \frac{2N_0\Delta}{\alpha_k S_1 n_{qp}},
\end{equation}
where $\alpha_k$ is the kinetic inductance fraction, $N_0$ is the single-spin density of states at the Fermi level ($1.72 \times 10^{10}\,\, \mu \mathrm{m}^{-3}\mathrm{eV}^{-1}$ for aluminum~\cite{Gao2008c}), $n_{qp}$ is the quasiparticle density, and
\begin{equation}
S_1 \approx \frac{2}{\pi}\sqrt{\frac{2\Delta}{\pi k_B T}}\sinh{\left(\frac{hf}{2k_B T} \right)}K_0 \left(\frac{hf}{2k_B T} \right),
\end{equation}
expresses the frequency and temperature dependence. 
Here $K_0$ is the modified Bessel function of the second kind.
When operating our detectors at 200~mK, $S_1$ ranges from about 0.075 to 0.15 across our readout band.
For these dark measurements, we can substitute the thermal quasiparticle density given by ~\cite{Gao2008c}
\begin{equation}
\label{eqn:nqpth}
n_{qp,\mathrm{thermal}} \approx 2N_0 \sqrt{2\pi k_B T \Delta} \,\exp{\left(-\frac{\Delta}{k_B T}\right)}
\end{equation}
into Equation~\ref{eqn:qimb} to obtain
\begin{equation}
Q_{i,MB} \approx \frac{\pi}{4\alpha_k}\frac{e^{\Delta/(k_B T)}}{\sinh{(\frac{hf}{2k_B T})}K_0(\frac{hf}{2k_B T})}.
\end{equation}

The loss due to TLS is represented by the product of a geometrical filling factor $F_{\mathrm{TLS}}$, and a loss tangent $\delta_{\mathrm{TLS}}$~\cite{Gao2008a}:
\begin{equation}
Q_{\mathrm{TLS}}^{-1}=F_{\mathrm{TLS}}\delta_{\mathrm{TLS}}.
\end{equation}
The TLS loss tangent depends on temperature and electric field as:
\begin{equation}
\delta_{\mathrm{TLS}} = \delta_0\tanh{\left(\frac{hf}{2k_B T}\right)}\left[\frac{1}{\sqrt{1+|E/E_c|^2}}\right],
\end{equation}
where $\delta_0$ is the loss tangent at zero temperature and zero field, $E$ is the electric field (which is related to readout power), and $E_c$ is the critical field for TLS saturation, defined by Gao et al.~\cite{Gao2007}
Since the measurements described in this section were all taken at a fixed readout power and the quality factor does not change appreciably below 300~mK where TLS effects are dominant, the electric field in the resonator $E$ should be roughly constant, so we can treat the electric field dependence term as a constant. 
Since this constant is guaranteed to be less than one, we simply take it to be equal to one to obtain lower limits on $\delta_{\mathrm{TLS}}$.
While $F_{\mathrm{TLS}}$ can be estimated from electromagnetic simulations, the measurements here cannot disentangle its value from $\delta_0$, so we treat the product as a single parameter in the fit.

The terms in the model for the frequency shift are the prediction from Mattis-Bardeen theory, the effect of TLS, and a constant offset $x_{\mathrm{offset}}$, which is added for convenience to take into account the fact that the reference frequency to which the fractional change is measured is arbitrary.
The prediction for the fractional frequency shift from Mattis-Bardeen theory is given by
%
%
\begin{equation}
x_{\mathrm{MB}} = -\frac{\alpha_k S_2}{4 N_0 \Delta}n_{qp},
\label{eqn:deltaf_mb}
\end{equation}
where
\begin{equation}
S_2 \approx 1+\sqrt{\frac{2\Delta}{\pi k_B T}}\exp{\left(-\frac{hf}{2k_B T} \right)}I_0 \left(\frac{hf}{2k_B T} \right).
\end{equation}
Here $I_0$ is the modified Bessel function of the first kind.
When operating our detectors at 200~mK, $S_2$ is approximately 3.8 across our readout band.
For these dark measurements we again substitute the equation for the thermal quasiparticle density from Equation~\ref{eqn:nqpth} to obtain
\begin{align}
x_{\mathrm{MB}} &\approx -\frac{\alpha_k}{4N_0\Delta} \bigg[1+\sqrt{\frac{2\Delta}{\pi k_B T}}\exp{\left(-\frac{hf}{2k_B T} \right)}I_0 \left(\frac{hf}{2k_B T} \right)\bigg] \nonumber \\
& \qquad{} \times \bigg[ 2N_0\sqrt{2\pi k_B T \Delta}e^{-\Delta/(k_B T)} \bigg] ,
\end{align}

The frequency shift induced by the temperature dependent TLS loss is:
\begin{align}
x_{\mathrm{TLS}} &= \frac{F_{\mathrm{TLS}}\delta_0}{\pi} \left[ \operatorname{Re} \left[\Psi{\left(\frac{1}{2}+\frac{hf}{2\pi j k_B T}\right)}\right] -\log \left(\frac{hf}{k_B T}\right) \right] \nonumber  \\
& \qquad{} \times \left[\frac{1}{\sqrt{1+|E/E_c|^2}}\right],
\end{align}
where $\Psi$ is the complex digamma function.
As before, we assume the electric field dependence term is equal to one and interpret the resulting $F_{\mathrm{TLS}}\delta_0$ as a lower limit.
Over the range of temperatures and readout frequencies we use, the term involving $\Psi$ is essentially constant and approximately equal to $\operatorname{Re}(\Psi(1/2))\approx-1.96$. 

%
%
%
%
%
%
In practice, there is a degeneracy between $\alpha_k$ and $\Delta$.
We first attempted to fix $\Delta=1.76k_B T_c$ using $T_c=1.46$~K as measured in Section~\ref{sec:film_properties}.
The resulting fits implied $\alpha_k\approx0.35$, which is inconsistent with the measured resonance frequencies and film properties.
Instead, we fixed $\alpha_k=0.65$ using those measurements and found that the fits (shown in Figure~\ref{fig:mb_fit}) required that $T_c\approx1.60$~K, or that $\Delta=1.93 k_B T_c$.
Similarly elevated $T_c$-to-$\Delta$ conversion factors have been suggested for aluminum resonators in the literature.~\cite{Janssen2014}

The free parameters in the fit are then $\Delta$, $F_{\mathrm{TLS}}\delta_0$, $Q_{i,loss}^{-1}$, and the nuisance parameter $x_{\mathrm{offset}}$.
We fit the data using three variations of the model, as shown in Figure~\ref{fig:mb_fit}.
First, we held $F_{\mathrm{TLS}}\delta_0 = 0$ and $Q_{i,loss}^{-1} = 0$ and fit only the data above 250~mK, where the response should be well described by the pure Mattis-Bardeen theory. 
The resulting fit, shown as a solid green line, does indeed describe the data above 250~mK well, but it offers no explanation of the limited $Q_i$ and non-monotonic ``back-bending" behavior seen at lower temperatures.
Next, we attempted to fit the full model including $F_{\mathrm{TLS}}\delta_0$ and $Q_{i,loss}^{-1}$.
The result is the dashed red line.
Here, the ``back-bending" in the frequency data dominates the fit, requiring a large value of $F_{\mathrm{TLS}}\delta_0 \approx 1.4 \times 10^{-4}$, which implies more loss than is actually seen. Thus $Q_{i,loss}^{-1}=0$ in this case.
While this model does show ``back-bending," the shape is not exactly in agreement with the data.
Others have reported similar discrepancies at low temperatures for TiN LEKIDs.~\cite{Swenson2013}
Previous studies of the temperature dependance of TLS have been made at much higher readout frequencies, where it is possible to probe the minimum in the $x_{TLS}$ equation that occurs at $T=hf/(2k_B)$.
For these devices, the temperature of this minimum is around 2.8~mK, which is not accessible with our cryostat.
Finally, we held $F_{\mathrm{TLS}}\delta_0 = 0$, and found that, aside from the ``back-bending," the limit to $Q_i$ could be explained as a constant loss of $Q_{i,loss}^{-1} \approx 1.0 \times 10^{-6}$.
This value is much more reasonable than the value implied for $\delta_0$ implied by the TLS fit, but of course cannot explain the ``back-bending."
This residual loss could potentially be attributed to a residual, constant population of quasiparticles.

%
%
%

\subsection{Optical Testing}


\subsubsection{Quasiparticle Lifetime}

The response of the resonances in the readout bandwidth is limited by the resonator ring-down time $\tau_r = Q_r / (\pi f_0) \sim 300~\mu \mathrm{s}$.
It is thus difficult to measure the quasiparticle lifetime $\tau_{qp}$ using these resonances.
We used a vector network analyzer to find higher-order resonances with lower quality factors so that the resonator bandwidth would be large enough to easily measure $\tau_{qp}$.
We targeted resonances with high enough quality factors such that the response to illumination was easily detectable.
We read out these resonances with an analog homodyne setup.

We tested a nine-element array from the same wafer as the twenty-element array. The array was mounted in a gold-plated copper package sealed with aluminum tape.
The detectors were illuminated through small holes in the tape by a 660~nm red LED coupled to a 2~mm diameter plastic fiber.
We studied a resonance with a loaded quality factor of 6300 and a resonant frequency of 810 MHz.

\begin{figure}[t]
\centering
\includegraphics[width=0.41\textwidth]{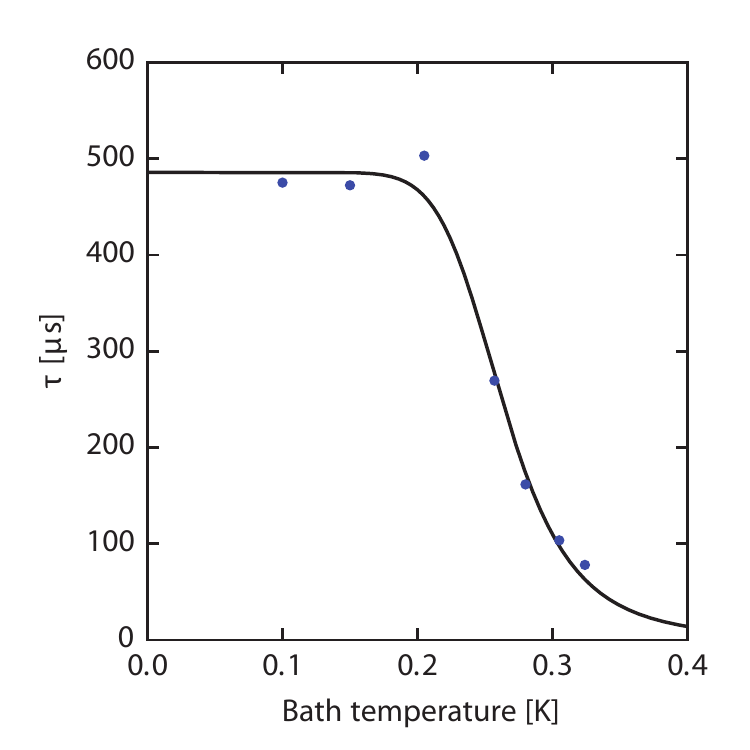}
\caption{{
Measurements of a detector time constant as a function of bath temperature, extracted from fits to the time-domain response of a higher-order resonance to an LED pulse.
Statistical error bars from the fitting process would be smaller than the data points.
The solid black curve is a fit of Equation~\ref{eqn:thermal_quasiparticle_lifetime} to the data, assuming a thermal quasiparticle density, and assuming $\Delta=1.76k_B T_c$.
The results are $\tau_{\mathrm{max}} = 488 \, \pm 16 \, \mu \mathrm{s}$ and $n^* = 363 \pm 38\, \mu \mathrm{m}^{-3}$.
%
If we instead assume $\Delta=1.93k_B T_c$, as implied by the Mattis-Bardeen fits in Section~\ref{sec:bath_sweeps}, $n^* = 160\,\pm20\,\mu \mathrm{m}^{-3}$.
}}
\label{fig:tau_qp_vs_bath_temperature}
\end{figure}

We fit an exponential decay to the time-domain response to an LED pulse and found that the response was fit well by a single time constant.
We measured this time constant as a function of bath temperature. As shown in Figure~\ref{fig:tau_qp_vs_bath_temperature}, the response time varied approximately as
\begin{equation}
\tau
  =
  \frac{\tau_{\mathrm{max}}}{1 + n_{qp, \mathrm{thermal}} / n^*},
\label{eqn:thermal_quasiparticle_lifetime}
\end{equation}
where $n_{qp, \mathrm{thermal}}$ is given by Equation~\ref{eqn:nqpth}. This is the expected behavior of the quasiparticle lifetime.~\cite{zmu}

\begin{figure}[t]
\centering
\includegraphics[width=0.95\columnwidth]{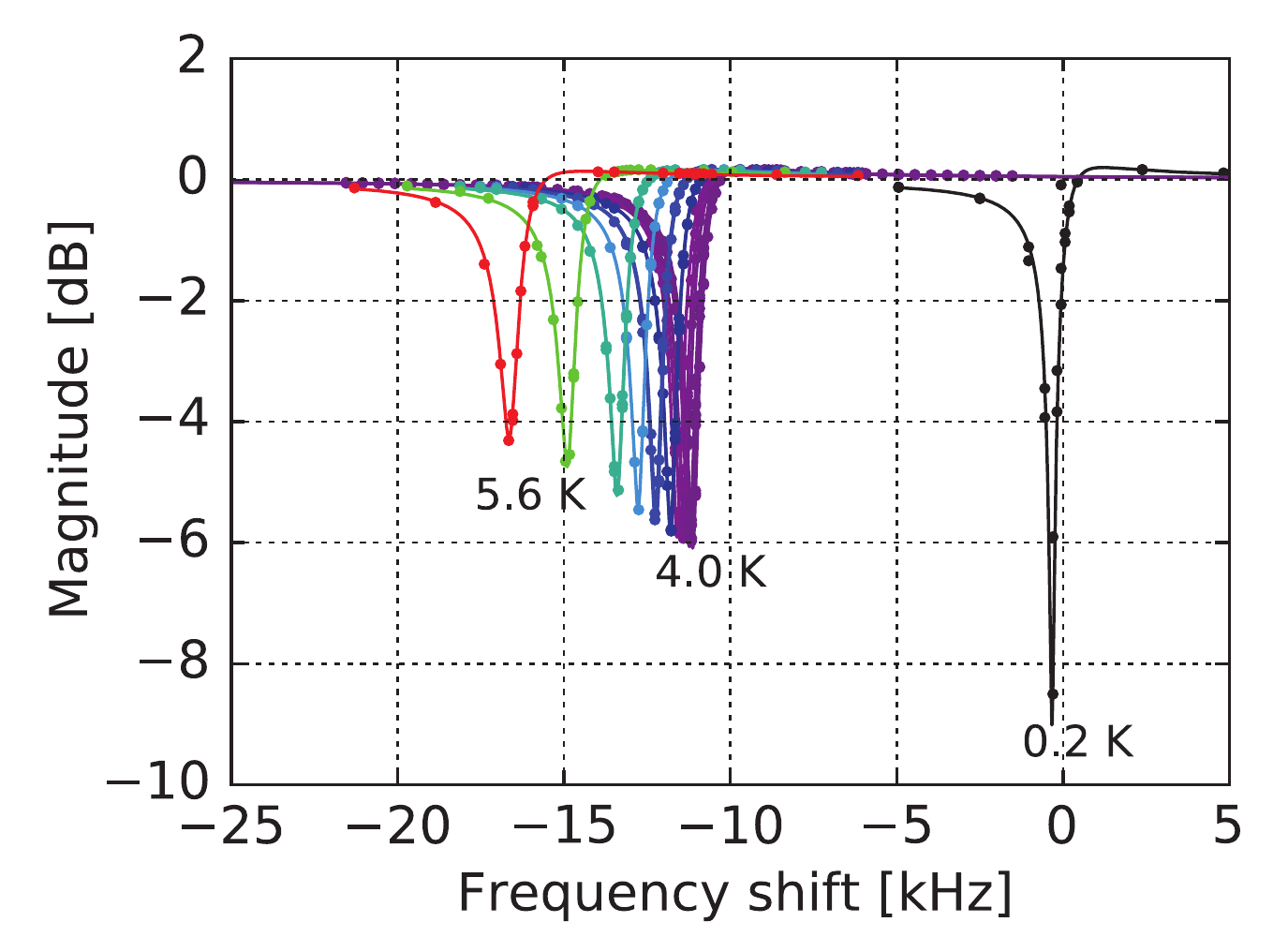}
\caption{{
Resonance sweeps of a single resonator with changing optical load. The dots are measured points and the lines are fits. The resonant frequency at 200~mK is 86.31195~MHz. For these measurements the probe power was $\sim-111$~dBm, and the bath temperature was 200~mK. The measurement labeled 0.2~K was taken in the dark configuration.
}}
\label{fig:tbb_sweeps}
\end{figure}

%

\begin{figure}[t]
\centering
\includegraphics[width=0.95\columnwidth]{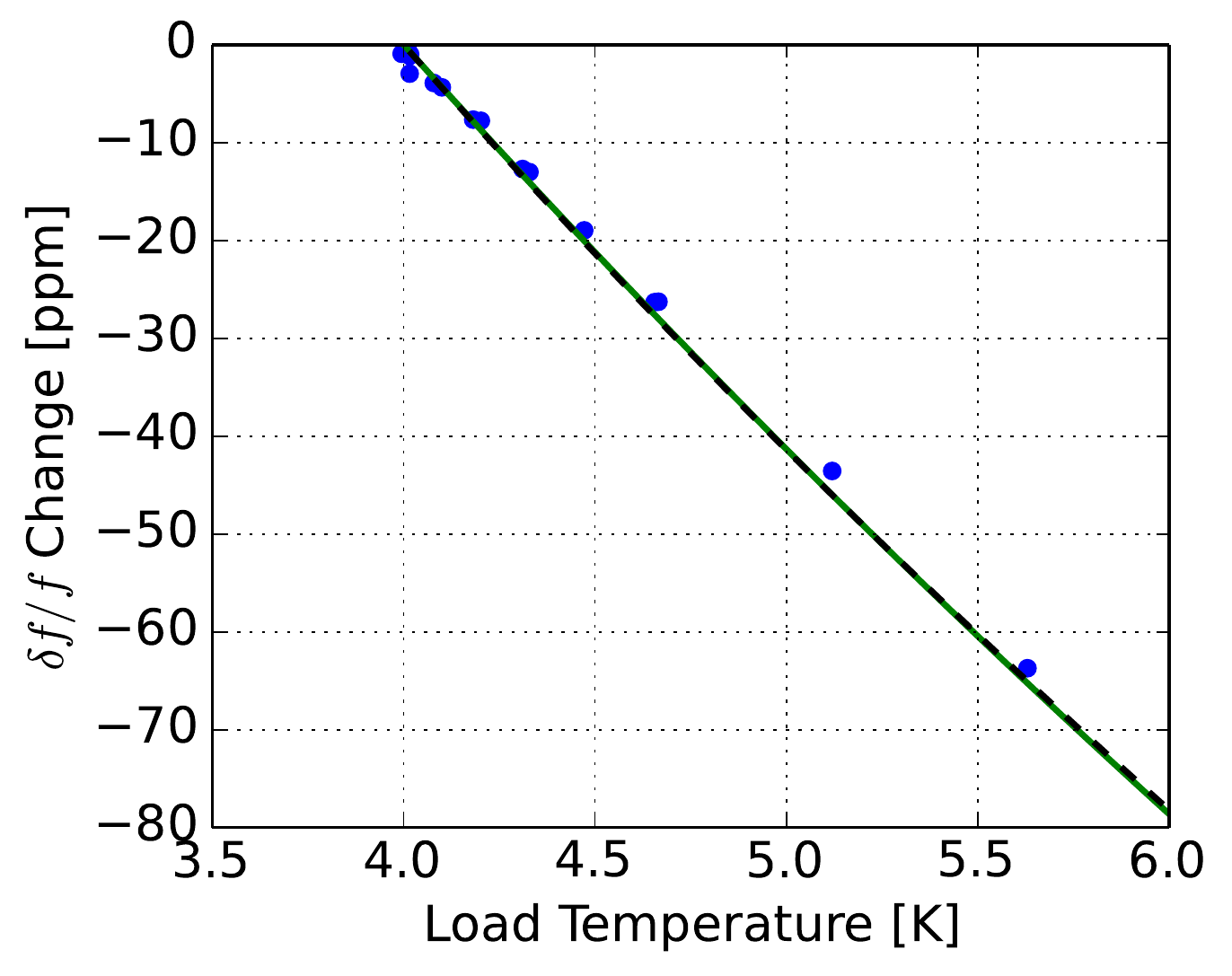}
\caption{{
Fitting the resonance curves shown in Figure~\ref{fig:tbb_sweeps} provides measurements of the resonant frequency as a function of blackbody temperature, plotted here as the fractional frequency change relative to the resonant frequency measured with the blackbody temperature at 4~K. 
Over the limited range of blackbody temperatures probed, the predicted response deviates by only a small amount from linear. 
The measured frequency response has a slope of approximately 40~ppm/K. 
This fractional responsivity is seen consistently across all resonators. 
The solid green line shows the expected response assuming  $\tau_{max}=500$~$\mu$s, and $n^{*}=400$~$\mu$m$^{-3}$, as suggested by the time constant measurements, and a total optical efficiency of $\eta=0.14$.
The dashed black line superimposed shows an alternative explanation for the data with the same value of $\tau_{max}$, but with $n^{*}=160$~$\mu$m$^{-3}$, which is closer to the typical value reported in the literature.~\cite{zmu}
In this case, $\eta=0.32$.
}}
\label{fig:responsivity}
\end{figure}


\subsubsection{Responsivity}

The temperature of the blackbody load, $T_{bb}$, is changed to measure the responsivity of the device to optically produced quasiparticles. In practice, $T_{bb}$ ranged from 4 to 6~K, or approximately 1.7 to 3~pW for one mode with two polarizations over the 120 to 180~GHz band. 
The detectors are designed to be predominantly sensitive to a single polarization with an absorption efficiency $\eta_{1} = 0.72$. The orthogonal polarization is expected to have $\eta_{2} = 0.13$. Since our optical source is unpolarized, the predicted power absorbed by the detectors is calculated using a single mode with two polarizations, and the appropriate optical efficiencies are applied to each polarization. We refer to the total optical efficiency as $\eta$. For use in a polarimeter, there would be a polarization selective element, such as a wire grid, before the devices. 

The measured response of a resonator as a function of optical loading is shown in Figures~\ref{fig:tbb_sweeps} and~\ref{fig:responsivity}. The resonant frequency response is linear across the range of optical powers tested, and is about 40~ppm/K in all the devices.   

If thermal quasiparticles are negligible, the quasiparticle density is~\cite{McKenney2012}
\begin{equation}
n_{qp}
  =
  n^*
  \sqrt{1 + \frac{2 \eta_{pb} \eta P_o \tau_{\mathrm{max}}}{\Delta V_L n^*}}
  - n^*,
  \label{eqn:nqpoptical}
\end{equation}
where $n^*$ is the film-dependent characteristic quasiparticle density,~\cite{zmu} $\eta_{pb}$ is the conversion efficiency (assumed to be $\sim0.7$ over our band~\cite{Guruswamy2014}), $P_o$ is the incident optical power, $\tau_{\mathrm{max}}$ is the maximum quasiparticle lifetime, $h$ is the Planck constant, $\nu$ is the photon frequency, and $V_L$ is the inductor volume.
The incident optical power in a waveguide from a blackbody of temperature $T_{bb}$ is 
\begin{equation}
P_o 
 =
  \int_{\nu_{l}}^{\nu_{h}} \frac{2\,h\,\nu^{3}}{c^{2}\,(e^{h\,\nu/k\,T_{bb}}-1)} \frac{n}{\lambda^{2}} \mathrm{d}\nu 
\end{equation}

where n is the number of dual-polarization modes, $\lambda$ is the incident wavelength, and $\nu_{l}$ and $\nu_{h}$ are  the low and high frequency edges of the spectral band. In this experiment the spectral band is defined by the waveguide cutoff at 130~GHz and the low-pass filter at 170~GHz.  

The absorbed power is $P = P_o \eta$, where $\eta$ is the absorption efficiency of the device. 
The frequency shift caused by an optical load $P_o$ can be computed by substituting Equation~\ref{eqn:nqpoptical} into Equation~\ref{eqn:deltaf_mb} giving

\begin{eqnarray}
x
 &=&
  -\frac{\alpha_k S_2}{4 N_0 \Delta}
  \left[
  n^* \sqrt{1 + \frac{2 \eta_{pb} \eta P_o \tau_{\text{max}}}{\Delta V_L n^*}} - n^*
  \right]
\label{eqn:optical_freq_shift}
\end{eqnarray}

We observe a linear relationship between $T_{bb}$ and the frequency shift as seen in Figure~\ref{fig:responsivity}.
Note that for the experimental setup $T_{bb}$ is very nearly proportional to $P_o$.
The range of optical powers over which we measure the responsivity is small and therefore it is difficult to distinguish a $\sqrt{P_o}$ from a $P_o$ dependence. 
A linear (or nearly linear) relationship has been observed by many groups and widely reported in the literature with no clear explanation of the phenomenon for both aluminum~\cite{Gao2008c} and TiN devices. ~\cite{Noroozian2012b,Hubmayr2014}
These other measurements tested over a wider range of optical power. 
By taking the partial derivative of Equation~\ref{eqn:optical_freq_shift} with respect to $P_o$, we can compute the expected responsivity:
\begin{equation}
\frac{\partial x}{\partial P_o} = \frac{\alpha_k S_2 \tau_{max} \eta_{pb} \eta}{4 N_0 \Delta^{2} V_L} 
	\bigg[ 1 + \frac{2 \eta_{pb} \eta P_o \tau_{max}}{\Delta V_L n^{*}} \bigg]^{-1/2}.
\end{equation}

%
Substituting in $\alpha_k=0.65$, $S_2=3.8$, $V_L=1870$~$\mu\mathrm{m}^3$, $\eta_{pb}=0.7$, $\eta = 0.14$, $\tau_{max}=500$~$\mu$s, $n^*=400 \, \mu \mathrm{m}^{-3}$, and $\nu=150$~GHz yields $\partial x/\partial P_o \approx 45$~ppm/pW. For a blackbody load temperature of 4~K, we expect $P_o \approx1.75$~pW and $\delta P_o / \delta T_{bb} \approx 0.88$~pW/K, which implies $\partial x/\partial T \approx 40$~ppm/K, in good agreement with our measurements.

Alternatively, if $n^*=160 \, \mu \mathrm{m}^{-3}$, as would be the case if $\Delta=1.92k_B T_c$ (see Figure~\ref{fig:tau_qp_vs_bath_temperature}), $\eta = 0.32$ would also yield a responsivity of 40~ppm/K.

\subsubsection{Optical Versus Thermal Response}

\begin{figure}[t]
\centering
\includegraphics[width=0.95\columnwidth]{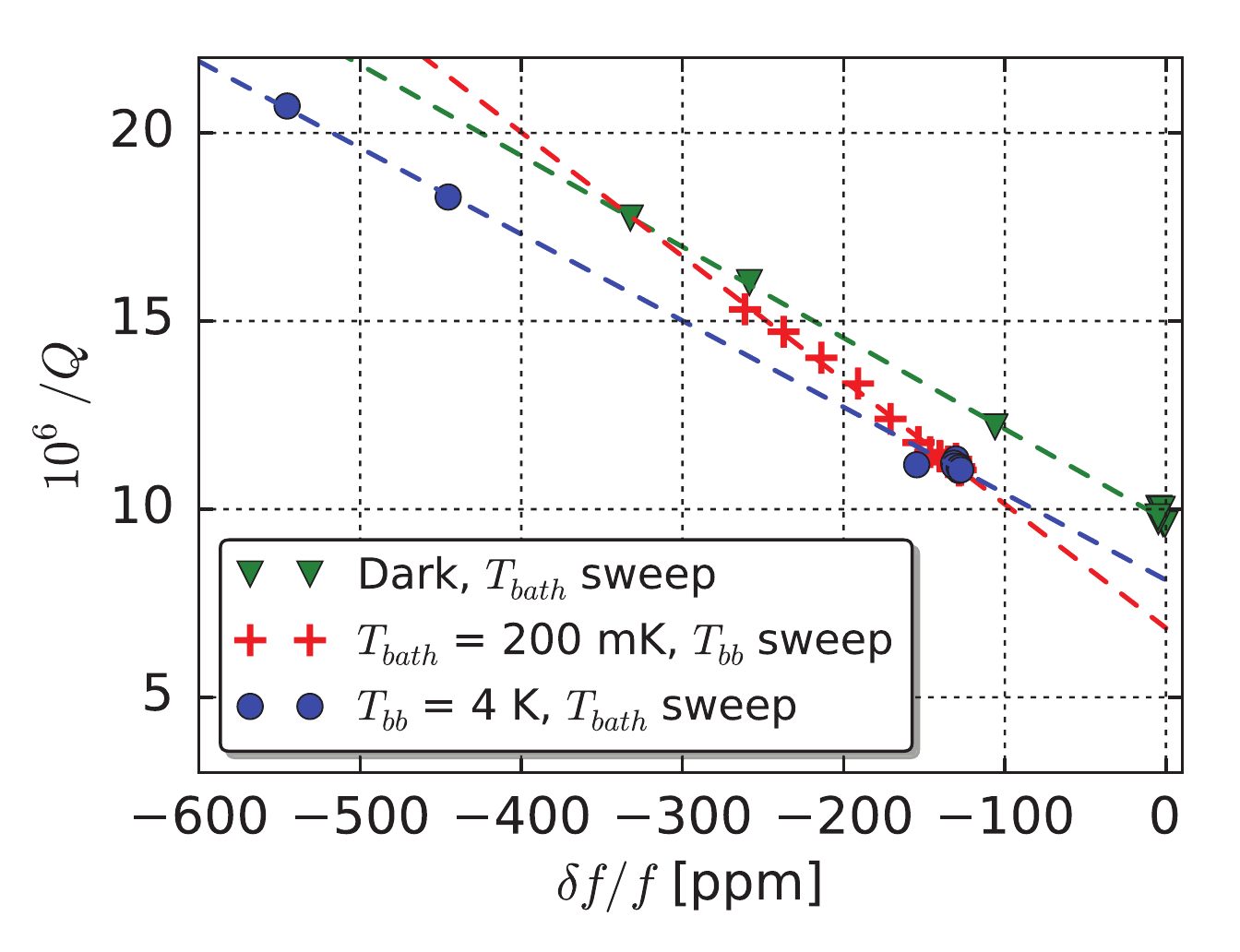}
\caption{{
Resonant frequency shift plotted versus inverse quality factor for a single resonator, showing good agreement to the linear behavior predicted by Mattis-Bardeen theory.
The points derived from bath temperature sweeps (blue dots and green triangles) show a consistent slope, which differs from that found when sweeping the blackbody load temperature (red crosses).
The dashed lines are linear fits to the data, used to derive the $\beta$ parameters plotted in Figure~\ref{fig:betas}.
}}
\label{fig:qivsf0}
\end{figure}

\begin{figure}[t]
\centering
\includegraphics[width=0.95\columnwidth]{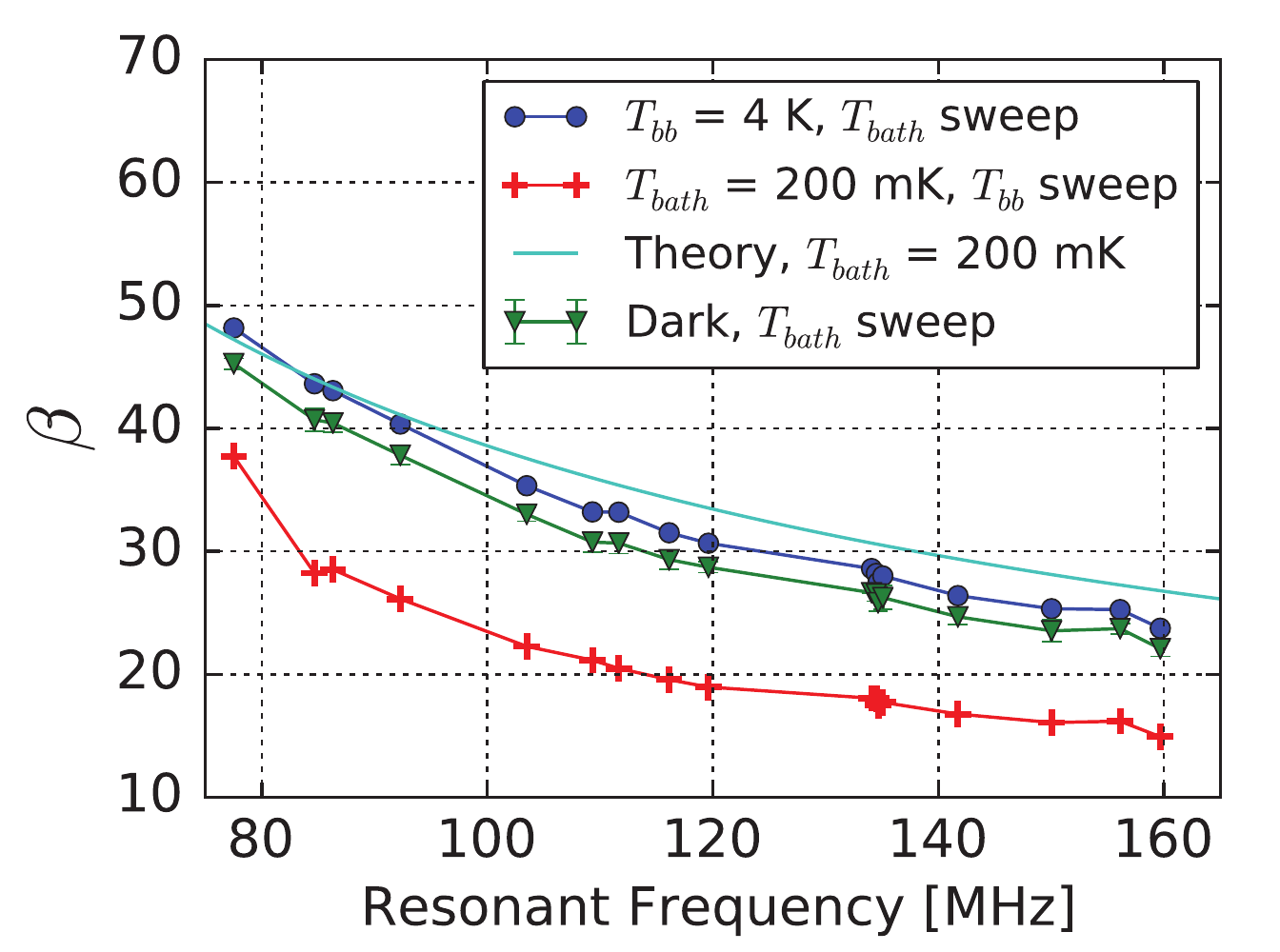}
\caption{{
Ratio of frequency responsivity over dissipation responsivity ($\beta$) plotted versus resonant frequency for all resonators.
The thick line shows the theoretical curve expected for a thermal quasiparticle distribution, given by Equation~\ref{eqn:beta}.
The errors on all measurements are similar to those shown on the dark, $T_{bath}$ sweep curve. 
The marker symbols and colors match the corresponding data in Figure~\ref{fig:qivsf0}, from which the values of $\beta$ are derived.
}}
\label{fig:betas}
\end{figure}

A comparison of the response to optical power and to changing bath temperature is presented in Figure~\ref{fig:qivsf0}.
Mattis-Bardeen theory predicts that for a given change in quasiparticle density, the fractional change in resonant frequency should be very nearly linearly related to the change in the quality factor.~\cite{Gao2008c}
This behavior is evident in these devices.
The slope of this linear relationship is the ratio of the frequency responsivity to the dissipation responsivity, and is defined in the literature as~\cite{zmu}:
\begin{equation}
\label{eqn:beta}
\beta
  =
  -\frac{2\delta f_0 / f_0}{\delta Q_r^{-1}}
  =
  \frac{S_2}{S_1}
  =
  \frac{1 + \sqrt{\frac{2\Delta}{\pi k_B T}}
    \exp{\left(-\frac{hf}{2k_B T}\right)}
    I_0{\left(\frac{hf}{2k_B T}\right)}}
  {\frac{2}{\pi}\sqrt{\frac{2\Delta}{\pi k_B T}}
    \sinh{\left(\frac{hf}{2k_B T}\right)}
    K_0{\left(\frac{hf}{2k_B T}\right)}}.
\end{equation}
This quantity and the theoretical prediction are plotted in Figure~\ref{fig:betas}.

Gao et al.~\cite{Gao2008c} showed theoretically and experimentally that $\beta$ should be the same for quasiparticles generated thermally or optically.
Our measurements seem to show a different behavior.
We find that $\beta$ matches the theoretical prediction when changing the device temperature, either in a dark environment, or with a constant optical flux from the black body load held at a fixed temperature.
However, when the bath temperature is kept fixed and the black body load temperature is varied, $\beta$ is appreciably smaller.
%
This effect has been reported for other devices as well.~\cite{Janssen2014}
They suggest that this effect could be explained if the optical pair breaking is non-uniform across the inductor.
This could be the case for our devices: we designed them to efficiently couple to the circular waveguide, but electromagnetic simulations do show variations in the electric field of the incident millimeter wavelength radiation across the inductor.
%
%
%
We plan to further investigate this phenomenon in the future. 


\subsubsection{Noise}

Noise measurements were made by recording time series of the complex transmission at a fixed probe frequency. 
For each noise measurement, a frequency sweep was performed and fit to the resonator model given in Equation~\ref{eqn:resonator} to determine the optimal value of the probe tone frequency. 
When analyzing the noise data, the complex time series is scaled by the complex $dS_{21}/df$ vector determined from the resonator model. 
The real and imaginary components of the resulting scaled time series then correspond to the fluctuations in resonant frequency and dissipation, respectively. 
This also results in the units of the frequency fluctuation time stream being hertz. 
As a check, we also apply the eigenvector decomposition technique suggested by Gao et al.,~\cite{Gao2007} which reports a constant rotation angle between the two principal components of the fluctuation spectrum within the device bandwidth, thus confirming that the fluctuations can be decomposed using a simple rotation.
Initially, noise measurements were taken with the pulse-tube cooler on.
This did not produce significant low-frequency noise when the devices were in a copper package, as the resonator quality factors were limited by the coupling to the lossy normal metal.
However, when testing with the superconducting aluminum package, the larger $Q_{i}$ increased the responsivity, and noise from the pulse-tube cooler was clearly evident in the noise spectrum.
The spectra shown here were taken with the pulse-tube cooler off, while the ADR continued to regulate the temperature of the detector package.
Turning off the pulse-tube cooler is not a viable option for a deployed instrument, so we are working to better understand and mitigate the source of this extra noise.

\begin{figure}[t]
\centering
\includegraphics[width=0.9\columnwidth]{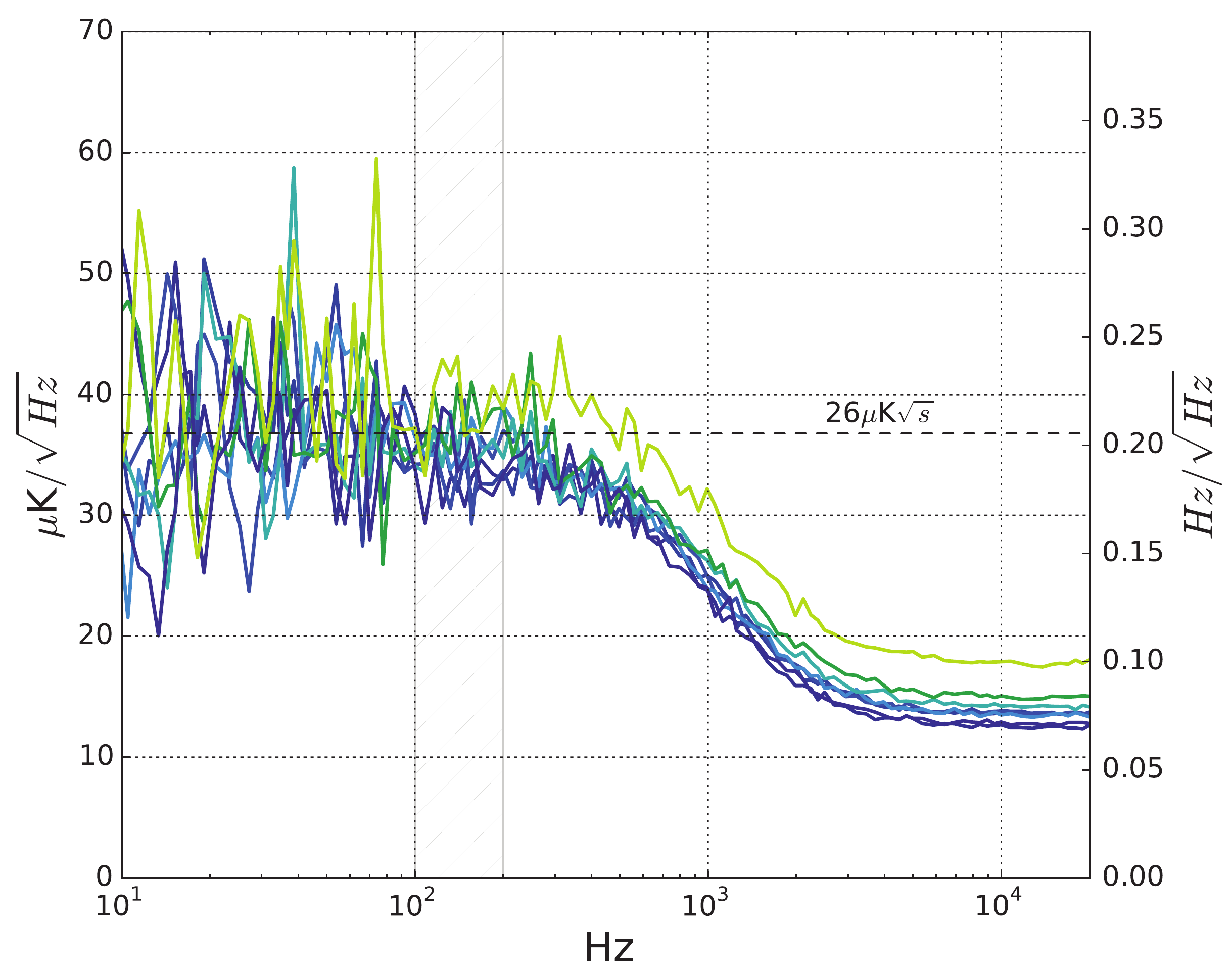}
\caption{{
The noise spectrum of a single, typical detector is shown at different optical loads.
The blackbody temperatures range from 4~K (purple) to 5.3~K (yellow), with the noise level increasing slightly with temperature. 
The average value in the region indicated near 100 Hz corresponds to the measurements shown in Figure~\ref{fig:net1}. 
The probe tone power was approximately -105~dBm.
}}
\label{fig:noisespectrum}
\end{figure}

\begin{figure}[t]
\centering
\includegraphics[width=0.9\columnwidth]{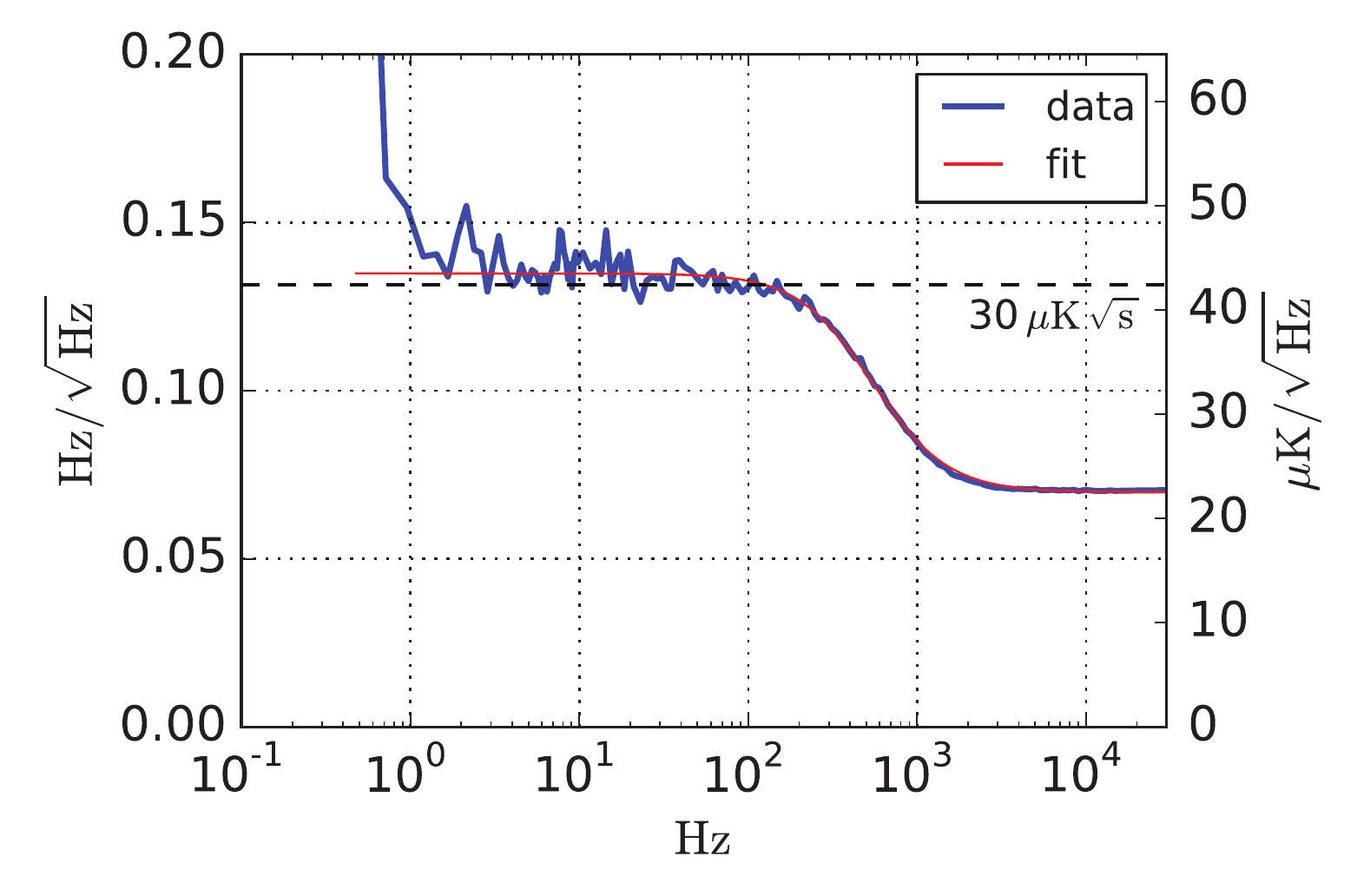}
\caption{{
The average spectrum (thick blue line) of a resonator obtained from 13 time series, each 30 seconds long. 
This is a different resonator than that shown in Figure~\ref{fig:noisespectrum}. 
Multiple time series were taken to see the noise down to very low frequencies, which appears to be very flat.
For modulation schemes, such as with a continuously rotating half-wave plate (HWP), that we envision for future CMB projects,~\cite{skip2013,Araujo2014} we are particularly interested in the noise performance between 10 and 50~Hz.
The thin red line is a fit of the spectrum to a Lorentzian model, showing good agreement.
The resonator ring-down causes the roll-off at $460$~Hz.
The steep rise at the lowest frequencies is due to drift of the blackbody load temperature.
The probe tone power for this measurement was approximately -113~dBm.
}}
\label{fig:average_noisespectrum}
\end{figure}

A series of noise spectra are show in Figure~\ref{fig:noisespectrum} for load temperatures between 4 and 5.3~K. 
Figure~\ref{fig:average_noisespectrum} shows the result of averaging 13 spectra taken in succession with a fixed 4~K load temperature and 200~mK bath temperature to better show the quality of the noise at low frequencies.
%
%
The measured $\mbox{Hz}/\sqrt{\mbox{Hz}}$ fluctuation spectrum is converted to $\mu\mbox{K}/\sqrt{\mbox{Hz}}$ units by multiplying by the slope of the measured frequency shift as a function of blackbody temperature, using the measurements shown in Figure~\ref{fig:responsivity}.
This can in turn be interpreted as noise equivalent temperature (NET) by dividing by $\sqrt{2}$~$\sqrt{\mbox{s}}/\sqrt{\mbox{Hz}}$.
The resonator in Figure~\ref{fig:average_noisespectrum} response rolls off at $\sim$460~Hz. This bandwidth corresponds well with the expected half-width at half-maximum (HWHM) resonator bandwidth of $(f_0/Q_r)/2$,
%
where, for this resonator, $f_0=92.277432$~MHz and $Q_r=99595$.
The high quality factor and low resonant frequency obscures the roll-off due to the quasiparticle lifetime $\tau_{qp}$. 

\begin{figure}[t]
\centering
\includegraphics[width=\columnwidth]{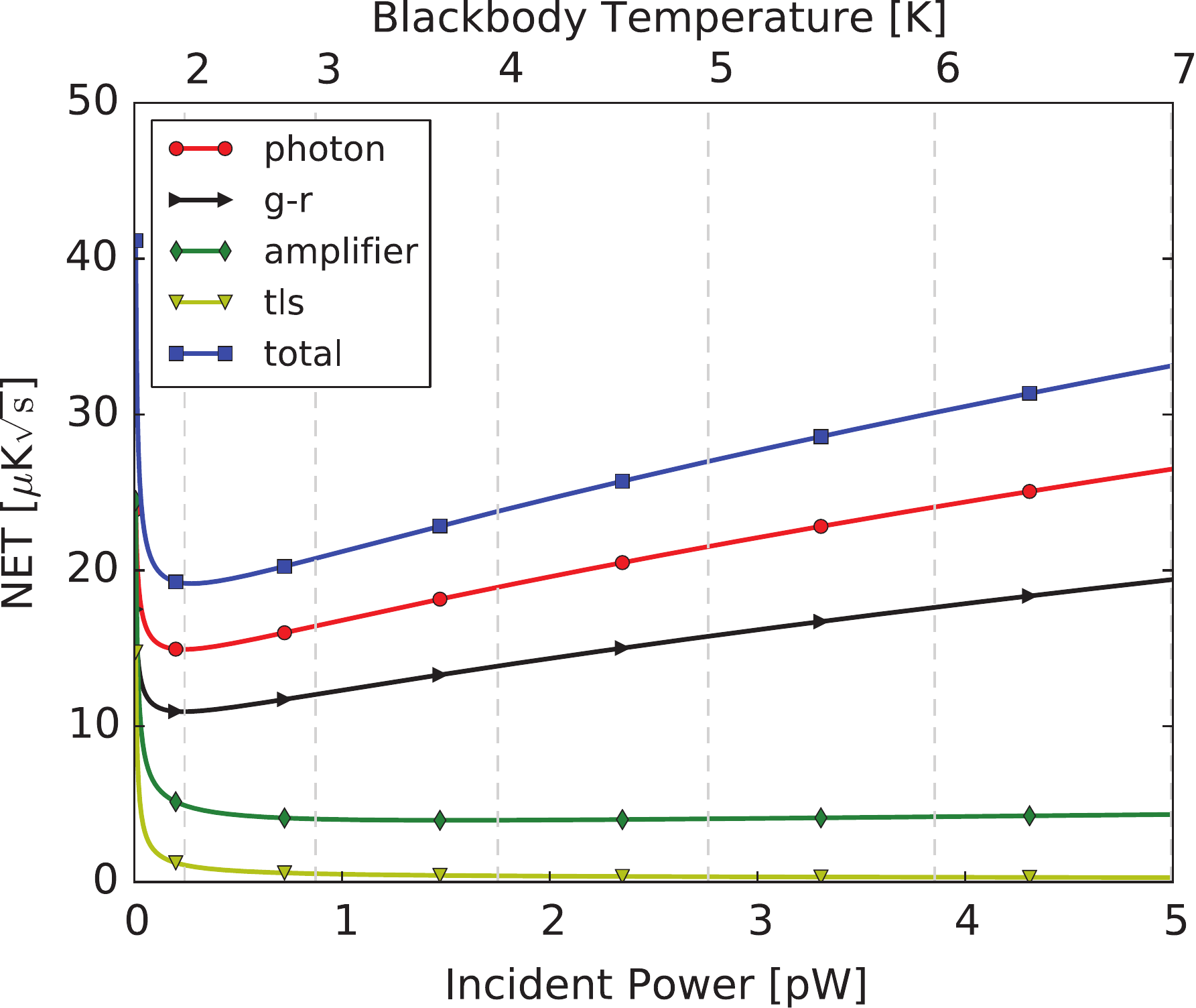}
\caption{{
The predicted levels for the different noise sources versus optical power. 
The total NET is composed of photon noise and detector noise (under load). 
The detector noise includes three components: g-r noise, readout amplifier noise, and TLS noise.
The expected TLS noise level is computed using a semi-empirical formula~\cite{Gao2008b} and fiducial scaling values were provided by MAKO.~\cite{McKenney2012}
The expected amplifier and TLS values are calculated assuming operation near bifurcation, which changes with load. 
The gray vertical lines correspond to blackbody load temperatures from 2 to 7~K in steps of 1~K.
}}
\label{fig:net1}
\end{figure}

Using the semi-empirical model, we predicted the noise equivalent power (NEP) contributions for the LEKIDs in our cryogenic set up and then converted them to NET as shown in Figure~\ref{fig:net1}.
The photon noise is calculated as 
$\mathrm{NEP}_{photon} = \sqrt{2 \eta P_{o} h \nu(1 + n_{o})}\,/\eta$,
where $ n_{o}$ is the photon occupancy number and is negligible for the range of powers tested.~\cite{devisser2014, zmu}
The absorption is expected to be high ($>$70$\%$) in the polarization for which the detectors were designed to be sensitive, which we refer to as $\eta_1$. 
Additional optical loading may possibly couple into the detectors from the orthogonal polarization ($\eta_2$) or leaked power from adjacent resonators ($\eta_{leakage}$).  
For the plot in Figure~\ref{fig:net1}, we used $\eta_{1} = 0.72$, $\eta_{2} = 0.13$, and $\eta_{leakage} = 0.075$, so $\eta = (\eta_1 + \eta_2)/2 + \eta_{leakage} = 0.5$.
%
%
The expected generation-recombination noise  can be approximated as $\mathrm{NEP}_{GR} \approx \sqrt{2 \eta P_{o} \Delta / \eta_{pb}}/\eta$, and should be the dominant detector noise source in the case of photon noise limited detectors.~\cite{yates} 
The NET is calculated using $\mathrm{NET} = \mathrm{NEP} / (\sqrt{2} \,dP_o(T)/dT)$.
We emphasize this is the NET at the load temperature in our experiment, not $\mathrm{NET_{CMB}}$.

%
%
%
%
The optical efficiency of the detectors influences both the expected photon and g-r noise levels. 
Due to the uncertainty of the optical efficiency, we predict a range of a expected NET values for the detectors. 
For a photon noise limited detector, the predicted total detector NET value is approximately 20 to 30~$\mu$K$\sqrt{\mbox{s}}$ at 2~pW of incident power.
This predicted NET range is for an optical efficiency range of $\eta =$0.3 to 0.7 for a single mode with two polarizations.
%
%

\begin{figure}[t]
\centering
\includegraphics[width=\columnwidth]{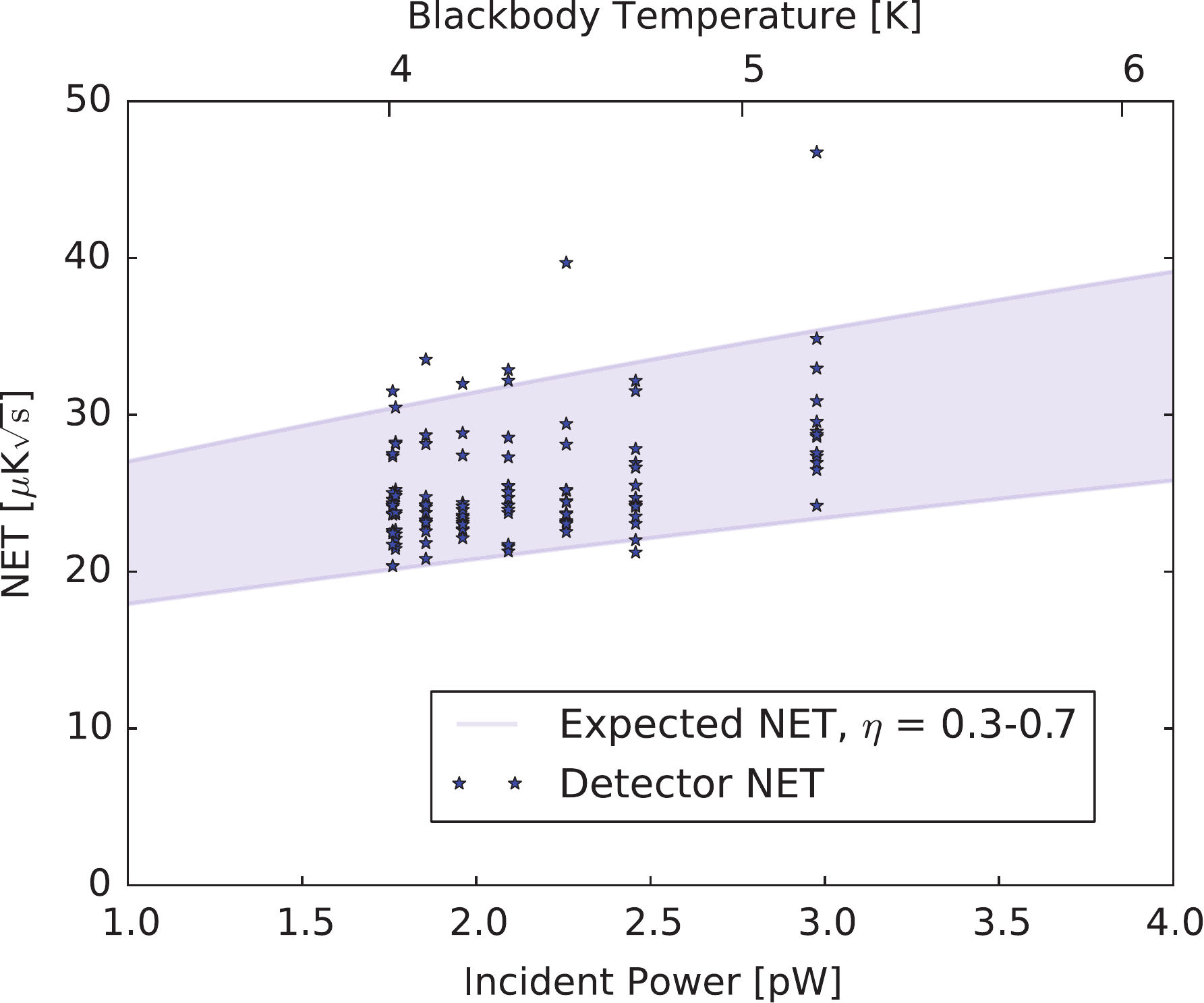}
\caption{{
Measured NET values and the predicted NET range. 
The incident power at the horn aperture is calculated from the blackbody temperature $T_{bb}$ using an emissivity of 0.92 for a single mode in two polarizations.
The absorption efficiency of the detectors can be simulated but is unmeasured.
The blue band shows the range of NETs expected for absorption efficiencies between $\eta = 0.3$ and $\eta = 0.7$. 
The stars show measured total NETs including the photon and detector noise. 
The data were measured with a constant probe tone power of approximately $-105$~dBm.
}}
\label{fig:net2}
\end{figure}

As seen in Figure~\ref{fig:net2}, the NET of the detectors on the tested array fall in the range 26$\thinspace\pm6$~$\mu$K$\sqrt{\mbox{s}}$ with a 4~K optical load.
We estimate the random error on individual NET measurements to be $\sim$10\%.
We do not find any systematic relationship between NET and resonant frequency.
Over the range of blackbody temperatures tested the noise remains fairly constant. However, as shown in Figure~\ref{fig:net2} the dependence of the noise on temperature is very shallow. The test setup was designed to directly measure the NET at 4~K using only small changes in load temperature. In future experiments, we will use another optical source to probe the noise over a wider range of optical power. 

The predicted amplifier noise contribution in units of $\mathrm{Hz}/\sqrt{\mathrm{Hz}}$ is~\cite{Lee2000}
\begin{equation}
e_{f,amp} = \sqrt{\frac{4k_B T_N}{P_{read}}}\frac{Q_c}{Q_r^2}f_0,
\end{equation}
where $T_N\sim4$~K is the noise temperature of the LNA and $P_{read}$ is the probe tone power.
While there is uncertainty in the exact value of $P_{read}$, the estimated value of $-105$~dBm predicts $e_{f,amp}\sim0.05$~$\mathrm{Hz}/\sqrt{\mathrm{Hz}}$ for $f_0=100$~MHz, in reasonable agreement with the measurements shown in Figure~\ref{fig:noisespectrum}.

Given the uncertainty in our exact optical absorption and pair breaking efficiencies, the measured noise data shown in Figure~\ref{fig:net2} is consistent with our devices having a contribution from TLS noise at a level ranging from negligible to approximately equal to the photon noise level. 
We find the noise spectrum is white down to the lowest frequencies measured, as seen in Figure~\ref{fig:average_noisespectrum}. This behavior has been reported by other groups,~\cite{superspec2014} however previous measurements of devices with TLS noise typically have a $e_{f, TLS} \sim\nu^{-0.25}$ shape, where $\nu$ is the frequency of the noise spectrum.~\cite{McKenney2012} 
%

\section{Conclusion}
\label{sec:conclusion}

In this paper we presented a LEKID design and a horn-coupling strategy that appears promising for cosmic microwave background (CMB) studies.
Our LEKIDs were made from a single thin aluminum film deposited on a silicon wafer and patterned using standard photolithographic techniques at STAR~Cryoelectronics.
We described the cryogenic testing apparatus and the testing program.
Finally, we presented the results from our optical testing, dark testing and aluminum film characterization measurements.
Our data were compared with Mattis-Bardeen theory for consistency.
These results show the multiplexing scheme works well, the yield across multiple LEKID arrays is 91\%, and the NETs are in the range 26$\thinspace\pm6$~$\mu \mbox{K} \sqrt{\mbox{s}}$.

Future work will focus on further decreasing the TLS effects, increasing the number of elements in the array, developing a dual-polarization design and better understanding the performance of these devices.
In particular, the following items warrant further investigation: (i) the response of the detectors is somewhat more linear than expected, (ii) the measured $\beta$ is different for thermal and optical quasiparticles, and (iii) the $T_{bath}$ sweeps in Figure~\ref{fig:mb_fit} show evidence for TLS effects but the relationship between $\delta_{\mathrm{TLS}}$ and the TLS noise level is unknown, in particular for low-frequency, aluminum LEKIDs. 
We also plan to measure the noise of these devices over a wider range of optical power using an improved optical source.

We have already fabricated a second-generation wafer that underwent an hydrofluoric acid dip during fabrication, and we expect the TLS effects of the devices on this wafer will be reduced.

A small fraction of the radiation that is not absorbed or reflected propagates laterally in the dielectric substrates and this signal could produce detector-to-detector cross-talk.
To mitigate  this effect, we will metalize the fused silica wafer with titanium nitride (TiN) between the horns and patterned to act as an efficient millimeter-wave absorber with an effective sheet resistance of approximately 150~$\Omega$. This TiN layer also helps to absorb ballistic phonons propagating in the silicon from energy deposited by cosmic rays.
We are working with STAR~Cryoelectronics on TiN films with the desired $T_{c}$.
This work could also lead naturally to developing commercial TiN LEKID designs sensitive to different frequencies. 

The present geometry of the inductor absorbs an average of less than 10\% of the cross polarization as predicted by electromagnetic simulations.
A rectangular waveguide or wire grid polarizer in front of the focal plane will further define the polarization selectivity of the focal plane, and make it truly single-polarization. We plan to add one of these varieties of polarization selectivity to future detector modules. 
%


\begin{acknowledgements}

This LEKID research is supported by a grant from the Research Initiatives for Science and Engineering (RISE) program at Columbia University to BRJ, and by internal Columbia University funding to AM. 
Matthew Underhill at the Micromachining Laboratory at Arizona State University machined the aluminum module with horns used for the measurements reported here. 
We thank the Xilinx University Program for their generous donation of FPGA hardware and software tools used in the readout system.
We also thank the anonymous reviewer for helpful comments.
This research has made use of NASA's Astrophysics Data System.

RC of STAR Cryoelectronics is a co-author and also owner of the foundry which fabricated the devices.
\end{acknowledgements}


\section*{Author Contributions}

All authors were involved in designing the detectors, interpreting the results, and editing the manuscript.
HM, DF, GJ, and BRJ wrote the manuscript and did the experimental measurements reported here.
RC oversaw the fabrication of the devices.
PA and CT provided the metal-mesh low pass filter.
VL wrote the initial version of the resonator fitting code. 
KB and DA contributed to the readout electronics.


\appendix
\section*{Appendix: Root of the bifurcation equation}
\label{sec:appendix}

The root of the cubic equation that appears in the bifurcation resonator model relevant to the branch of the bifurcation accessible by our measurements is given by:
\begin{equation}
y =  y_0/3 + \frac{y_0^2/9 - 1/12}{k_1} + k_1,
\end{equation}
where
\begin{equation}
k_1 = \sqrt[3]{a/8 + y_0/12 + k_2 + y_0^3/27} 
\end{equation}
and
\begin{equation}
k_2 = \sqrt{(y_0^3/27 + y_0/12 + a/8)^2 - (y_0^2/9 - 1/12)^3}.
\end{equation}
See Swenson et al.~\cite{Swenson2013} for a discussion of the branches of the bifurcation.


\bibliographystyle{aipnum4-1.bst}
\bibliography{paper.bib}


\end{document}